\documentstyle[psfig,astrobib,times]{mn-ab}

\def\spose#1{\hbox to 0pt{#1\hss}}
\def\la{\mathrel{\spose{\lower 3pt\hbox{$\sim$}}
        \raise 2.0pt\hbox{$<$}}}
\def\ga{\mathrel{\spose{\lower 3pt\hbox{$\sim$}}
        \raise 2.0pt\hbox{$>$}}}

\def\H0{{\it H}$_0$}
\def\Ms{{\it M}$_\odot$}

\def\q0{{\it q}$_0$}

\def\ergps{erg~s$^{-1}$}
\def\kmpspMpc{km~s$^{-1}$~Mpc$^{-1}$}
\def\Ms{{\it M}$_\odot$}

\def\Zs{$Z_{\odot}$}

\def\d{{\rm d}}
\def\ie{i.e.\ }
\def\eg{e.g.\ }
\def\esn{$\epsilon_{\rm SN}$}
\def\dpar#1#2{{\partial#1\over\partial#2}}
\def\vel{{\bf v}}
\def\stress{{\bf T}}
\def\half{{1\over2}}
\def\rhog{\rho_{\rm g}}

\def\figwidth{0.5\textwidth}

\title[Heating in the formation of X-ray clusters]
{Non-gravitational heating in the hierarchical formation
 of X-ray clusters}
\author[K. K. S. Wu et al.] 
{\parbox[]{6.5in} {K. K. S. Wu,$^{1,3}$ A. C. Fabian$^1$ and
P. E. J. Nulsen$^2$}\\
\\
$^1$ Institute of Astronomy, Madingley Road, Cambridge CB3 0HA, UK\\ 
$^2$ Department of Physics, University of Wollongong, Wollongong
     NSW 2522, Australia\\
$^3$ Physics Department, University of California, Santa Cruz, CA
95064, USA. E-mail: kwu@ucolick.org}
\date{}

\begin{document}

\maketitle

\begin{abstract}
  The strong deviation in the properties of X-ray clusters from simple
  scaling laws highlights the importance of non-gravitational heating
  and cooling processes in the evolution of proto-cluster gas. We
  investigate this from two directions: by finding the amount of
  `excess energy' required in intra-cluster gas in order to reproduce
  the observed X-ray cluster properties, and by studying
  the excess energies obtained from supernovae in a semi-analytic
  model of galaxy formation. Using the insights obtained from the
  model, we then critically discuss possible ways of achieving the
  high excess specific energies required in clusters. These include
  heating by supernovae and active galactic nuclei, the
  role of entropy, and the effect of removing gas through radiative
  cooling.
  
  Our model self-consistently follows the production of excess energy
  and its effect on gas halos. Excess energy is retained in the gas as
  gravitational, kinetic and/or thermal energy.  The density
  profile of a gas halo is then selected according to the total energy of
  the gas. Our principle assumption is that in the
  absence of non-gravitational processes, the total energy of the gas
  scales as the gravitational energy of the virialized halo---a
  self-similar scaling law motivated by hydrodynamic simulations. This
  relation is normalized by matching the model to the largest observed
  clusters.

  We model the gas distributions in halos by using a 2-parameter
  family of gas profiles. In order to study the sensitivity of results
  to the model, we investigate four contrasting ways of modifying gas
  profiles in the presence of excess energy.  In addition, we estimate
  the minimum excess energy required in a fiducial cluster of around 2
  keV in temperature by considering all available gas profiles. We
  conclude that the excess energies required lie roughly in the range
  1--3 keV/particle.
  
The observed metallicities of cluster gas suggests that it may be
possible for supernovae to provide all of the required excess
energy. However, we argue that this scenario is only marginally
acceptable and would lead to highly contrived models of galaxy
formation. On the other hand, more than enough energy may be available
from active galactic nuclei.
\end{abstract}

\begin{keywords}
galaxies: clusters: general -- galaxies: formation -- galaxies:
evolution -- cooling flows -- X-rays: galaxies
\end{keywords}

\section{Introduction} \label{introsec}

Much progress has been made in recent years in the modelling of galaxy
formation, partly in response to an unprecedented amount of new data,
especially for galaxies at high redshift.  This paper however aims to
constrain the model from the high-mass end, by tackling the properties
of X-ray clusters. This has the advantage that the results are
insensitive to the detailed physics of star formation and
feedback. Only a small fraction of the hot gas in clusters is able to
cool in a Hubble time, so that any star formation has little effect on
the structure of the gas halo. Since star formation and feedback are
two of the least understood components of galaxy formation, this seems
to be a natural approach to take.

On the other hand, X-ray clusters do contain a fossil record of the
complex star-formation history of their progenitors. The amount of gas
left in a cluster's halo depends on the amount consumed in processes
such as star formation. The heavy elements (or metals) observed in the
gas are the result of enrichment by supernovae over billions of years.
Like the metals, the energy injected into the gas by supernovae and active
galactic nuclei (AGN) is retained in the gas if it is not
radiated. We shall be particularly interested in this `excess energy'
that is retained in present day clusters. X-ray clusters
therefore provide important constraints on the history of a large
sample of baryons.

Broadly speaking, a complex physical system can be studied via
numerical methods, \eg N-body simulations, or via analytic
calculations. In galaxy formation theory, the {\em semi-analytic}
approach has come to refer to more than just an intermediate line of
attack, but to a specific class of models that use the hierarchical
merger tree as their starting point. In the Cold Dark Matter (CDM)
model \cite{bfpr84}, small halos virialize
first and progressively collapse into larger and larger halos. The
merger tree follows the masses of these halos as a function of time.
The evolution of the baryonic component in these halos---which
comprises $\sim 1/10$ of the total mass---receives a simplified yet
physical treatment that models processes such as cooling, star
formation and supernova feedback, to name a few.

Although N-body simulations of dark matter (DM) clustering now provide
perhaps the best understood piece in the jigsaw of how galaxies
formed, the evolution of the baryonic component remains much less well
understood. In both hydrodynamic+DM simulations and semi-analytic
models (SAMs), many of the above gas processes need to be
approximated by simple rules.  Nevertheless, using SAMs, we are able
to efficiently explore the unknown parameters in these processes, and
study the range of behaviour in these systems.  In this way, SAMs have
achieved notable success in modelling many properties of galaxies
\cite{wf91,kwg93,cafnz94,kc98,bcfl98,sp98,ghbm98}.

In this paper, we investigate the effect of excess energy on the
density profiles of gas halos, and thus on the properties of X-ray
clusters.  Excess energy is retained in the gas as
thermal, gravitational and/or kinetic energy as it passes through a
merger tree.  Even if the gas is ejected from a halo, it is expected
to recollapse into a larger halo at a later time, thus the excess
energy is not lost. As a first approximation, the excess energy in a
gas halo is given by the total energy obtained from non-gravitational
heating, minus the energy lost via radiative cooling. By
non-gravitational heating we refer to heating by sources such as
supernovae and AGN. The total energy released by such sources (though
not necessarily injected into or retained by the gas) comes to
several keV per particle when averaged over all baryons in the
universe.  It therefore has the potential to strongly influence the
properties of X-ray clusters and galaxies.

It has been known for some time that the match between theoretical
predictions and the observed properties of X-ray clusters is
significantly improved if we assume that the gas is `pre-heated' in
some way \cite{kaise91}. Hydrodynamic simulations without
non-gravitational heating or cooling \cite{nfw95,bn98} obtain X-ray
clusters that are approximately `self-similar', in the sense
that small clusters (with temperatures $\approx 2\times 10^{7}$ K) are
similar to large clusters ($T\sim 10^8$ K) scaled down in size. (Note
that densities do not change in such a scaling.)  However, the gas
halos of observed clusters are not self-similar.  For example, the
X-ray luminosities of small clusters are an order of magnitude less
than those predicted by scaling down the luminosities of large
clusters in this way.  This suggests that the gas distributions of
small clusters are less concentrated than in large clusters.  In order
to break the self-similarity of X-ray clusters, excess energy is
generally required. Excess energy affects small clusters much more
than large ones. It can make the gas distribution more
extended, or even remove some gas from the halo. Different models for
heating clusters and breaking their self-similarity have been studied
by a number of authors
\cite{kaise91,eh91,me94,nfw95,cmt97,wfn98,pcn99,bbp99,loewe99,pen99}.

In order to model the effect of excess energy on gas halos, it is
necessary to have a continuous range of gas profiles to choose from.
The gas profile with density proportional to $r^{-2}$ has been used
successfully in many SAMs to model galaxies. However, it is too simple
for modelling the properties of X-ray clusters.  In particular, the
core of the gas density profile has to be flattened in order to obtain
results that match the data (Wu, Fabian \& Nulsen 1998;
WFN98)\nocite{wfn98}.  In WFN98 we introduced a family of isothermal
gas profiles into our SAM. We assumed the gas to be in hydrostatic
equilibrium inside potential wells given by Navarro, Frenk \& White
(1997; NFW97)\nocite{nfw97} density profiles.  This family of gas
profiles enabled us to increase the temperature of a gas halo
uniformly, according to the excess energy in the gas. The main results
from that paper are that we were able to fit the observed properties
of X-ray clusters, including their gas fractions, metallicities, X-ray
luminosity-temperature relation, temperature function, X-ray
luminosity function and mass-deposition-rate function, by including
excess energies of $\sim 1$ keV/particle.

However, for a given total energy possessed by the gas halo (the sum
of its thermal and potential energies), the isothermal profile
represents only one solution out of a continuous range of possible
solutions. Furthermore, it is uncertain how heating modifies a gas
halo, since that 
depends on details of how the heating occurred. We therefore need to
test the sensitivity of results to the way that we modify the gas halo
when excess energy is present.  To do this, we extend the family of
isothermal profiles by requiring that gas halos obey polytropic
equations of state: $P\propto \rho_{\rm g}^\gamma$, where $P$ is
pressure and $\rho_{\rm g}$ is gas density.  Thus for a given
potential well and total gas mass, the gas profile has two
degrees of freedom, given by the parameter $\gamma$, which effectively
specifies the shape of the temperature profile, and
the normalization of the temperature profile. The isothermal profiles
are retrieved when $\gamma=1$, while progressively steeper temperature
gradients are obtained by increasing $\gamma$.  We thus have the
choice of increasing the temperature of a gas halo uniformly with
radius or preferentially towards the centre, depending on the `heating
model' that is used. One of the main purposes of this paper
is to constrain the level of excess energy that intra-cluster gas must
have in order to match the observed properties of X-ray clusters. We
then critically discuss possible ways of obtaining this level of
heating.

The SAM used in this paper is based on that described by Nulsen \&
Fabian (1997, 1995; NF97 and NF95)\nocite{nf97,nf95}.  A discussion of
the main areas of difference with other SAMs is given in NF97.
However, our study of X-ray clusters is not affected by such
differences, as their X-ray properties depend almost entirely on their
gas profiles only.

We use an open cosmology with $\Omega_m=0.3$ and no cosmological
constant.  A Hubble parameter of $H_0=50$ \kmpspMpc\ is assumed
throughout. We assume that density fluctuations are described by a CDM
power spectrum with a primordial spectral index of $n=1$ and
normalized to give $\sigma_8=0.75$.  In addition, we assume a baryon
density parameter of $\Omega_b=0.02 h^{-2}$ (where $H_0=100h$
\kmpspMpc) based on big-bang nucleosynthesis and deuterium abundance
measurements \cite{bt98,bntt99}. For $h=0.5$, this implies
$\Omega_b=0.08$ and an initial gas fraction of
$\Omega_b/\Omega_m=0.27$.

\subsection{Plan of the paper}

The main results from our model are discussed in
sections~\ref{clustersec} and \ref{galsec}. 

Section~\ref{clustersec} investigates the excess energies required in
X-ray clusters and the relevant parts of the model are described in
sections~\ref{treesec}, \ref{nfwsec} and \ref{gassec}.

Section~\ref{galsec} discusses the amount of excess energy obtainable
from supernova heating in our model and therefore requires knowledge
of our star formation model as described in the rest of
section~\ref{briefsec}.

In section~\ref{ensec} we discuss some effects not accounted for by
our model that may possibly contribute to the excess energy.
In the process, we give a more formal definition of excess
energy and discuss the theory behind the concept in some detail.

Finally, in section~\ref{breaksec} we discuss four possible scenarios
for breaking the self-similarity of clusters, aiming to be as
model-independent as possible. We consider three sources of
energy: supernovae, AGN and preferential removal of gas by cooling.
We also discuss the role of entropy in this problem
(section~\ref{ponmansec}) and emphasize that both energy and entropy
are important in determining the final gas distribution. 
In section~\ref{summary} we summarize our conclusions.

\section{Brief Description of the Model} \label{briefsec}
We begin with a general description of our model which can be applied
to any reasonable gas and DM halo profiles. More detailed discussions
of the gas processes and galaxy formation model
can be found in NF95 and NF97, which assumed
essentially the same physics as used here. In Appendix A we apply the
rules given in this section to the set of density profiles that we
shall adopt.

\subsection{Merger trees} \label{treesec}
Merger trees of virialized halos are simulated using the Cole \&
Kaiser (1988\nocite{ck88}) block model. In a `complete' simulation, we
use 20 levels of collapse hierarchy where the smallest regions are
$1.5\times 10^{10}$\Ms\ in mass. In the block model, masses increase
by factors of 2 between levels, so that the mass of the largest block
is $7.9\times 10^{15}$\Ms. This allows us to simulate the full range
of structures from dwarf galaxies to the largest present-day clusters.
However, if we are considering X-ray cluster properties only, it is
$\sim 1000$ times faster to simulate only the top 10 levels of the
collapse tree. The mass of the smallest regions is then $2^{10}\times
1.5\times 10^{10}=1.5\times 10^{13}$\Ms. In such low-resolution
simulations some additional assumptions need to be made, such as the
value of the gas fraction left over from the formation of galaxies.

Since every collapse of a block (which corresponds to a major infall
or merger) at least doubles the mass of the largest progenitor halo, a
new halo is said to virialize with each collapse. The virial radius,
$r_{200}$, is defined such that the mean density within it is 200
times the background density of an Einstein-de Sitter universe of the
same age.  The total mass of the halo inside $r_{200}$ is equal to the
mass of the collapsed block. Likewise, the gas mass inside $r_{200}$
is the contribution from the entire block (unless the excess energy is
so high that the gas halo is unbound). The new halo is given gas
and DM density profiles, which allow the estimation of basic
quantities such as the cooling time of the gas. 
From this starting point, the model proceeds to estimate the rate of
star formation, supernova feedback, metal enrichment and other
quantities that can be compared with observations. At the next merger,
the properties of the progenitor halos (\eg the mass of gas remaining) are
then incorporated into the new halo.

A collapse which is
followed too closely by a larger-scale collapse does not have time to
form a virialized halo. It is therefore not counted as a separate
collapse. We allow a minimum time interval between collapses which is
parametrized as a multiple of the dynamical time. Our results are not
sensitive to this parameter and it is given a value of 1.

\subsection{Cold and hot collapses} \label{coldhot}
For a gas halo to be
considered hydrostatic, the gas at any radius has to remain still for
at least the time it takes for sound to travel to the centre, which
can itself be approximated by the free-fall time. As discussed in
NF95, if the ratio of cooling time to free-fall time to the
centre, $\tau=t_{\rm cool}/t_{\rm ff}$, is less than $\sim 1$, then
the gas cools fast enough that it is not hydrostatically supported. It
fragments and collects into cold clouds which we assume to form stars
with a standard or slightly modified initial mass function (IMF). We
refer to this as a cold collapse and the gas that takes part in it as
cold gas.

When $\tau\ga 1$, a hydrostatic atmosphere of hot gas (at roughly the
virial temperature) is able to form.  In this case, a cooling flow
occurs if some gas has time to cool before the next collapse. Cooling
gas flows inward subsonically and remains hydrostatically supported.
In clusters of galaxies, cooling flows are common and observations
show that the gas that cools does not form stars with a standard IMF
but must remain as very small, cold clouds or form low-mass stars
\cite{fabia94}.  We refer to the product simply as baryonic dark
matter (BDM). A possible mechanism for the formation of low-mass stars
in cooling flows is described by \citeN{mb99}, for the case of
elliptical galaxies.

To estimate the masses of hot and cold gas produced in a collapse, we
use the gas and total density profiles to estimate $t_{\rm cool}$ and
$t_{\rm ff}$ as functions of radius. To simplify computation, $t_{\rm
  ff}$ is estimated using the free-fall time of a test particle in a
uniform background density, \ie $t_{\rm ff}=\sqrt{3\pi/16G\rho}$,
where $G$ is the gravitational constant and the total density at the
radius concerned is substituted for $\rho$. (This gives a slight
overestimate of $t_{\rm ff}$, as density actually increases towards
the centre.) We 
thus obtain $\tau(r)$, and compare it to a critical value, $\tau_0$,
to determine if gas is hot or cold. In
well-behaved cases $\tau$ increases monotonically with radius, so that
there exists a unique radius $r_{\rm cf}$, inside
of which gas is labelled as cold, outside
as hot.  As halo mass increases, the trend is for $r_{\rm cf}$ to move
from outside the virial radius to the centre. In other words, cold
collapse gives way to hot collapse as we go to more massive halos.  This
transition is quite abrupt and takes place over about one decade in
mass.

From the above, it is clear that no single gas profile can always describe
the gas halo. Cooling modifies the gas distribution, and in a cold
collapse the assumption of hydrostatic equilibrium breaks down
completely. However, the gas profile used in the model is only {\em
  notional}---defined as that obtained in a notional collapse with
cooling ignored (Nulsen, Barcons \& Fabian 1998)\nocite{nbf98}. Used
in this way, it allows us to estimate the behaviour of different
subsets of gas.  In the case of hot halos, if the part that has cooled
is small compared to the whole, then the density and temperature of
gas away from the cooled region do not change significantly as the
halo reestablishes hydrostatic equilibrium. The original gas profile
therefore gives reasonable estimates of bulk properties.

\subsubsection{The criterion when excess energy is large}
If the excess energy from heating is large enough to be comparable to
the binding energy of the gas halo (as defined in
section~\ref{gassec}), then $\tau$ may not increase monotonically with
radius (see Appendix A for examples). Such cases can
account for a fair fraction of low-mass galaxies because
of their smaller binding energies. This raises the question of whether
gas with $\tau<\tau_0$ {\em outside} a core of gas where $\tau>\tau_0$
still ends up cold after collapse.  Since the value of $\tau$ and its
interpretation are approximate, we opt for a simple criterion in such
cases, which determines whether all or none of the gas halo takes part
in a cooling flow (Appendix A).
We note that $\tau(r)$ is a fairly flat function of radius if the
strongly heated gas halo is isothermal.

\subsection{Star formation, supernova feedback, and cooling flows}
\label{sfsec}
Star formation is presumed to proceed rapidly in cold gas and leads
quickly to type II supernovae.  This is assumed to continue until the
energy from supernovae is sufficient to eject the remaining gas in the
halo to infinity, or until the cold gas is used up. If the gas halo is
not ejected, supernova energy can modify the gas
density profile by increasing the total energy of the remaining gas
(see section~\ref{gassec}). The effect of this is generally small but is
included for consistency. The remaining gas, which is hot, may then
take part in a cooling flow, depositing BDM if it manages to cool by
the next collapse or the present day. For halos which contain only hot
gas or cold gas, the situation is naturally
simpler than described.

We only follow the production of type II supernovae (SNII) in our
model.  Precise knowledge of the IMF is not required, since we only need
to know the number of SNII resulting from a certain amount of star
formation. It is generally assumed that the progenitors of SNII are
stars of mass $M>8$\Ms. For a standard IMF (more precisely the
Miller-Scalo IMF), we adopt the estimate of one SNII for every 80\Ms\ 
of stars formed with $M\leq 1$\Ms\ (Thomas \& Fabian
1990\nocite{tf90}).  In the simulations, we make the simplification
that stars with $M>1M_\odot$ are instantaneously recycled, so that
only the total mass of stars with $M\leq 1$\Ms\ is recorded. This
allows us to calculate the mass of stars remaining in
present-day clusters.  Since the lifetime of a star is approximately
$10^{10} (M/M_\odot)^{-3}$ years, the recorded stellar mass is a good
approximation of this quantity.

[The above suggests that the amount of gas in a halo could be
overestimated by the model, since in reality, stars of intermediate
mass ($1M_\odot<M<8M_\odot$) recycle their gas as planetary nebulae on
intermediate time scales.  However, we find that in newly-formed
halos, the stellar mass is almost always $\la 1/10$ of the gas mass,
so that the effect of recycled gas on the latter is small (halos of a
few $10^{12}$\Ms\ are an exception, as $\sim 1/3$ of them have more
stars than this). Another minor problem occurs when only a small
fraction of the gas in a halo is cold, so that most of the cold gas forms
stars.  In this case, the assumption of instantaneous recycling can
cause the amount of star formation to be overestimated (in the extreme,
all of the cold gas can be converted into stars with $M<1M_\odot$).
Fortunately, the fraction of stars formed in such situations is very small,
so that the error in the stellar mass of present-day clusters is less
than 1 per cent.]

In the simulations, we follow NF97 by boosting the above supernova
rate by a factor of 5.  Hence each SNII is associated with 16\Ms\ of
stars formed with $M<1M_\odot$. This corresponds to using a flatter
slope for the IMF. Since the bulk of star formation in our model
occurs as massive bursts in dwarf galaxies, it should not be
surprising to find that the IMF is modified under such circumstances.

To give an actual example, a power-law IMF with a slope of $x=0.9$
(the Salpeter IMF has $x=1.35$), and lower and upper cutoffs of
0.1\Ms\ and 50\Ms, gives 1 SNII for every 15\Ms\ of stars with
$M<1M_\odot$.  (Results are not very sensitive to the upper cutoff,
because very massive stars are rare.) Using this IMF, we can estimate
the error in our assumption that the stellar mass of a present-day
cluster is given by the stars with $M\leq 1$\Ms. Suppose the stars in
the cluster have an age of 5 Gyr instead of 10 Gyr, then the surviving
stars would be given by $M<0.5^{-1/3}=1.26M_\odot$. For the above IMF,
the stellar mass in the range $0.1M_\odot<M<1.26 M_\odot$ is 11
per cent greater than that in the range $0.1M_\odot<M<1M_\odot$.
The stellar mass of model clusters are therefore correct to $\sim 10$
per cent.

The energy per supernova available for the ejection of gas is
parametrized as $4\times 10^{50}$\esn\ erg \cite{spitz78}. Although
the total energy released by a supernova is typically $\sim 10^{51}$
erg, a large fraction of this is likely to be radiated, especially if
the supernova explodes in cold gas.  Each SNII is assumed to release
an average of 0.07\Ms\ of iron (Renzini et al.\ 1993\nocite{rcdp93}).
The solar iron abundance is taken to be 0.002 by mass (Allen
1976\nocite{allen76bka}). Renzini et al.\ find that the average iron
yield is fairly insensitive to the slope $x$ of the IMF.  We note that
a more recent compilation of average iron yields from a range of SNII
models \cite{ns98} shows a wider dispersion, ranging from 0.07 to
0.14\Ms\ of iron per SNII. However, most of these SNII models assume
that the progenitor stars have solar metallicity, whereas the bulk of
star formation in our model occurs in low metallicity dwarf galaxies.
If we only consider the low metallicity SNII models, then the range
narrows to about 0.07--0.09\Ms\ of iron per SNII.

When a new halo collapses, the mean iron abundance and mean excess
specific energy ($E_{\rm excess}$) of the gas are calculated and
assigned to the gas halo.  The excess energy of a gas halo, as it name
implies, is the increase in its total energy (defined below) relative
to the total energy it would have in the absence of any
non-gravitational processes. In the model it is approximated by the
total energy injected by supernovae, minus the energy radiated in
progenitor halos, over the history of the gas. If some gas is removed
from the gas halo by a cooling flow, $E_{\rm excess}$ is assumed to
stay the same for the remaining hot gas. The reduction in
$E_{\rm excess}$ by radiative cooling is easily accounted for, since
the cooled gas is either converted to stars/BDM, or is ejected from the halo
by supernovae. Since we assume that the gas is always ejected at the
escape velocity of the halo, the resulting value of $E_{\rm excess}$
is simply given by the binding energy of the gas halo (as defined
below).  Other mechanisms that may affect $E_{\rm excess}$ but are not
accounted for by the model are discussed in section~\ref{ensec}. In particular,
if the gas is displaced by strong heating, this can lead to an extra
`gravitational contribution' to the excess energy, that is usually
positive. Unfortunately, this contribution is in general difficult to compute
(without hydrodynamic simulations) and is likely to be model-dependent as well.
(The approximation to $E_{\rm excess}$ made by the model may be compared
to the approximation made when inferring the clustering of mass from
the clustering of galaxies, which is traditionally handled by a `bias
parameter'.)

\section{The distribution of total density in halos} \label{nfwsec}
We begin by specifying the total density profile of a halo, which
allows us to derive the shape of the potential well. This is then used
in the following section to derive gas density profiles.

From a series of N-body simulations in different cosmologies and
with both CDM and power-law fluctuation spectra, Navarro, Frenk \& White
(1997\nocite{nfw97}; NFW) found that the density profiles of
virialized halos obey a universal form, given by
\begin{equation}
  \rho(r)={\delta_c \rho_{\rm crit}\over (r/r_s)(1+r/r_s)^2},
\end{equation}
where $\rho_{\rm crit}=3H^2/8\pi G$ and $H$ is the Hubble parameter at
the time of collapse.  The characteristic density $\delta_c$ is
calculated according to a prescription described in the Appendix of
NFW.  This method amounts to setting the scale density, $\rho_s=\delta_c
\rho_{\rm crit}$, equal to 3000 times the background density when the
halo was `assembled', subject to an appropriate definition of this
assembly time. The assembly time is a function of halo mass and
redshift of virialization only (given the cosmology and fluctuation
spectrum).

From the value of $\delta_c$ and the mean density of the halo within $r_{\rm
  200}$, the scale radius $r_s$ is uniquely determined. Thus
$\delta_c$ is the only `degree of freedom' in the profile. For
convenience, $x=r/r_s$ is often used to denote radius. The value of
$x$ at the virial radius, $c=r_{\rm 200}/r_s$, is an important
parameter known as the concentration.

[On a technical point, our model actually differs slightly from
the original NFW prescription. This is because NFW defined the mean
density of a halo to be $200\rho_{\rm crit}$, whereas we have chosen
to follow the spherical collapse model more closely when calculating
the mean density. By following their prescription for calculating
$\delta_c$, we have preserved their explanation for its origin.
However, quantities such as $r_s$ and $c$ will differ slightly.]

\begin{figure}
\centerline{\psfig{figure=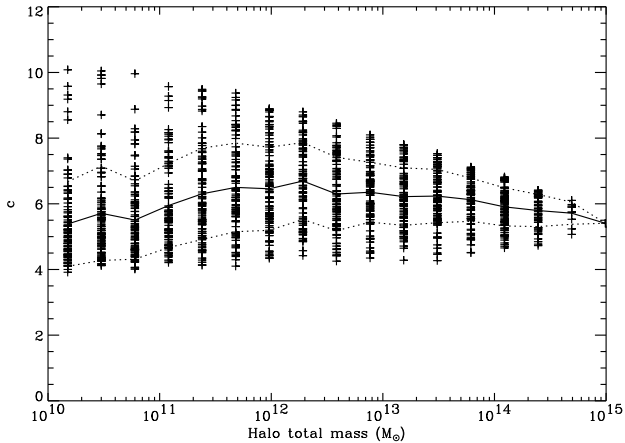,width=\figwidth}}
\caption[Scatter plot of concentration $c$ vs.\ halo mass]
  {Scatter plot of concentration $c$ vs.\ halo mass. Halo masses take
  discrete values in the block model. Each mass bin contains a maximum of 100
  points. Halos were selected randomly from the simulation regardless
  of redshift. The solid line gives the mean value of each mass bin,
  and the dotted lines are plotted one standard deviation from the
  mean.}
\label{figcofm}
\end{figure}

We make the further approximation that the NFW profile describes the
total density in a halo (\ie including the gas density)
and that it is truncated to zero for $r>r_{200}$. This
allows us to derive the gravitational potential as a function of $x$:
\begin{equation} \label{eqphi}
     \phi(x) = \alpha\left( -\frac{\ln(1+x)}{x} + \frac{1}{1+c} \right),
\end{equation}
where $\alpha = 4\pi G \rho_s r_s^2$. 

To illustrate typical values of $c$ obtained in this model,
Figure~\ref{figcofm} shows a scatter plot of $c$ against halo mass for
our choice of cosmology and fluctuation spectrum. For halos that
collapse at a given redshift, $c$ increases substantially with
decreasing mass, \eg the steep upper edge of the distribution is given
by halos that virialize at $z=0$.  However, for a given mass bin, $c$
decreases with increasing redshift. As a result, the mean value of $c$
does not vary much with halo mass, because less massive halos are more
likely to collapse at higher redshift.

\section{THE DISTRIBUTION OF GAS IN HALOS} \label{gassec}

Given the NFW potential well (\ref{eqphi}) and the total gas mass
within $r_{\rm 200}$, we make two further assumptions in order to
calculate the gas density profile. The first is that the gas is in
hydrostatic equilibrium, \ie
\begin{equation} \label{hstat}
  {\d P\over\d r}=-\rho_{\rm g}{\d\phi\over\d r},
\end{equation}
and the second is that $P$ and $\rho_{\rm g}$ are related by some
equation of state. For example, if we assume a perfect gas law and
isothermality, then $P\propto \rho_{\rm g}$ and the only parameter is
the temperature, $T$.  Once $T$ is specified, the gas profile is
uniquely determined. Below, we first describe the general
procedure that we use to determine such parameters.

We refer to the gas profile obtained in the absence of excess energy
as the {\em default profile}. Since the NFW profile is not
self-similar (see Fig.~\ref{figcofm}), it is not possible to define a
self-similar default profile for gas halos. In the absence of heating,
it is common to assume that $\langle T\rangle$ is proportional to
$\langle\sigma^2\rangle$ for the DM halo, where $\sigma$ is the
velocity dispersion and the brackets denote some form of average.
However, $\langle\sigma^2\rangle$ is non-trivial to compute for the DM
halo, and we are more interested with the total energy of the gas halo
than just its thermal energy, since any excess energy would be added
to the former. In order to retain some level of self-similarity,
we therefore postulate that the total specific energy of the gas halo,
$E_{\rm gas}$, is proportional to the specific gravitational energy of
the whole halo (which is modelled by the NFW profile):
\begin{equation}
  \label{energy}
    E_{\rm gas} =
   K \frac{1}{M_{\rm tot}} \int\frac{1}{2}\rho_{\rm tot}\phi\, \d V,
\end{equation}
where $E_{\rm gas}$ is defined by
\begin{equation}
  \label{egas}
  E_{\rm gas} \equiv \frac{1}{M_{\rm gas}} 
  \int\rho_{\rm g} \left(\frac{3kT}{2\mu m_{\rm H}}+\phi\right)\, \d V.
\end{equation}
The above integrals are performed out to the radius $r_{\rm 200}$,
$M_{\rm tot}$ and $M_{\rm gas}$ are the total mass and total gas mass
respectively, and the total density $\rho_{\rm tot}$ is given by NFW
density profile. The Boltzmann constant is denoted by $k$ and $\mu
m_{\rm H}$ is the mean mass per particle of the gas.  Note that
$E_{\rm gas}<0$ in order for the gas halo to remain gravitationally
bound.

The constant of proportionality $K$ is a parameter of the model. It is
calibrated by requiring that the default profiles of the largest
clusters approximate well those from X-ray observations.  We match to
the largest observed clusters because if heating does occur, we expect
it to have least effect on them.  Once $E_{\rm gas}$ is computed from
(\ref{energy}), the gas profile is uniquely determined if it is
selected from a family with only one parameter (\eg the isothermal
family).  In general, a numerical procedure is required to search for
the gas profile with the matching value of $E_{\rm gas}$.  The X-ray
clusters obtained in this way do closely follow the self-similar
scaling relations for $M_{\rm tot}$, $T$ and $L_{\rm X}$.

We refer to the value of $|E_{\rm gas}|$ given by (\ref{energy}) 
as the {\em binding energy}. As its name implies,
the binding energy is the excess specific energy required to unbind
the gas halo.

When the excess specific energy, $E_{\rm excess}$, is non-zero,
$E_{\rm gas}$ is increased accordingly:
\begin{equation}
    E_{\rm gas} =
   K \frac{1}{M_{\rm tot}} \int\frac{1}{2}\rho_{\rm tot}\phi\, \d V
                   + E_{\rm excess}.
\end{equation}
This allows the {\em heated} gas profile to be found.  In general, as
$E_{\rm excess}$ increases, the gas temperature increases and the gas
distribution becomes more extended (\ie the density profile becomes
flatter). Thus the excess energy goes into increasing both the thermal
energy and the potential energy of the gas halo.

If an isothermal family of gas profiles is used, heating increases the
temperature uniformly with radius.  Frequently, properties such as the
luminosity of an X-ray cluster or the amount of gas able to cool in a
given time are sensitive only to the gas density near the centre.
Therefore, if we increase the temperature preferentially towards the
centre, then we can obtain the same changes in these properties for
less excess energy. A convenient way of modelling non-isothermal
profiles is to use a polytropic equation of state: $P\propto \rho_{\rm
  g}^\gamma$. There are then two degrees of freedom, represented by
$\gamma$ and the constant of proportionality in the polytropic
equation.  Since there are two parameters, a continuous range of gas
profiles now have the same value of $E_{\rm gas}$. Thus a further
constraint is required to determine the gas profile uniquely.

A {\em heating model} is obtained by specifying
\begin{itemize}
\item[a)] the constraint used to determine the default profile, and
\item[b)] the path in parameter space followed by the gas profile as
  $E_{\rm excess}$ increases.
\end{itemize}
In order to obtain a good match to the largest clusters, the parameter
$K$ is allowed to depend on a). Thus, $K$ may also be regarded as part
of the heating model. The specification of a heating model is of
course artificial; in reality, the gas profile is determined by
additional factors such as the gas entropy distribution and how
shock heating occurs.  In lieu of a more complex model, we shall use a
few contrasting heating models to test the sensitivity our results.

\subsection{A 2-parameter family of gas density profiles} \label{familysec}

We now derive the family of gas profiles used in our model, assuming a
polytropic equation of state and a perfect gas law. If we first let
$\gamma=1$, then $T$ is constant and equation~\ref{hstat} gives
\begin{equation}
  \rho_{\rm g}(r) \propto \exp\left( -\frac{\mu m_{\rm H}}{kT} \phi(r) \right).
\end{equation}
Inserting the expression (\ref{eqphi}) for the NFW potential yields
\begin{equation}
  \label{isod}
  \rho_{\rm g}(r) \propto (1+x)^{\eta/x},
\end{equation}
where $\eta=\mu m_{\rm H}\alpha/(kT)$ is a dimensionless parameter
that characterizes the slope of the density profile.  Recall that
$\alpha$ is the characteristic gravitational potential of the NFW
profile. The mean value of $\eta$ obtained by fitting this model to
highly luminous X-ray clusters is approximately 10 \cite{ef98}.

For $\gamma\neq 1$, we use $P\propto \rho_{\rm g}^\gamma$ to eliminate
$P$ in equation~\ref{hstat}, and then use $\rho_{\rm
  g}^{\gamma-1}\propto T$ to get
\begin{equation}
  {\d\over\d r}\left(kT\over\mu m_{\rm H}\right) = 
                 -{\gamma-1\over\gamma}{\d\phi\over\d r}.
\end{equation}
Substituting for the potential gives
\begin{equation}
  \label{polyt}
  {T\over T_{200}} = 1+{\gamma-1\over\gamma}\eta_{200}
                       \left({\ln(1+x)\over x}-{\ln(1+c)\over c}\right),
\end{equation}
where $\eta_{200}=\mu m_{\rm H}\alpha/(kT_{200})$ is the value of
$\eta$ at the virial radius (where $x=c$). Thus, using $\gamma>1$ causes the
temperature to increase monotonically towards the centre.  Substituting
$\rho_{\rm g}\propto T^{(1/\gamma-1)}$, we get
\begin{equation}
  \label{polyd}
  {\rho_{\rm g}\over\rho_{\rm g,200}}=
    \left[1+{\gamma-1\over\gamma}\eta_{200}
      \left({\ln(1+x)\over x}-{\ln(1+c)\over c}\right)\right]^{1\over\gamma-1},
\end{equation}
where $\rho_{\rm g,200}$ is the gas density at the virial radius. It is
straightforward to show that this approaches the isothermal form
(\ref{isod}) as $\gamma\rightarrow 1$. We henceforth use the
parameters $\gamma$ and $\eta_{200}$ to specify the gas profile.

It is also useful to compute the `entropy', $s=T/n_e^{2/3}$, where
$n_e$ is the electron density, and $n_e\propto\rho_{\rm g}$. For our
purposes, $s$ may simply be regarded as a label for the adiabat that the
gas is on. For the gas to be stable to convection, the entropy must
increase with radius.  When $\gamma=5/3$, the entropy is constant with
radius, thus the atmosphere is marginally stable to convection.
Atmospheres with higher values of $\gamma$ and steeper temperature
gradients convect to reduce the temperature gradient. Hence 5/3 is the
maximum value of $\gamma$ used in the model. The minimum value used is
$\gamma=1$. We do not use lower values of $\gamma$ as there is
little evidence for the temperature in halos to increase with radius,
both from X-ray cluster observations and hydrodynamic simulations.

\begin{figure}
\centerline{\psfig{figure=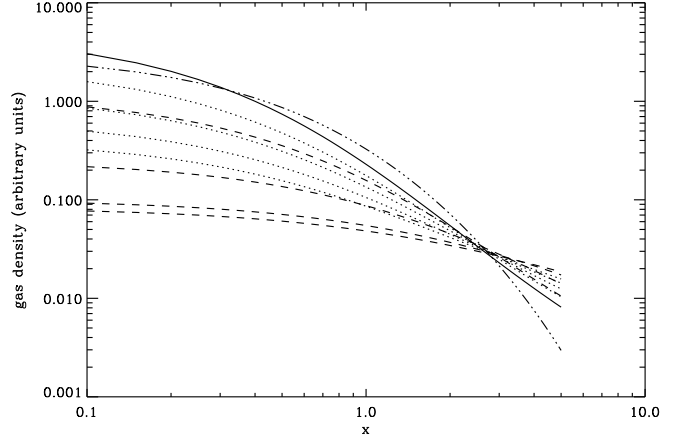,width=\figwidth}}
\caption[Examples of model gas density profiles]
{Gas density profiles, with parameters representative of those
  obtained in heating models A and B (see Fig.~\ref{figmodel} and
  text). The same total gas mass and NFW potential well were used
  throughout (we set $c=5$).  The solid curve (the default profile for
  the purposes of this figure) uses $\eta_{200}=10$ and $\gamma=1$.
  The series of dotted profiles have $\gamma=1$ but decreasing values
  of $\eta_{200}$: namely $\eta_{200}=8.5$, 7.1, 6.0 and 5.1. These
  values were chosen so that their total specific energies, $E_{\rm
    gas}$, increase at regular intervals.  The flattest profile,
  $\eta_{200}=5.1$, has zero total energy and is marginally bound.
  The series of dashed curves have the same total energies as the
  dotted curves, but have the following parameters:
  $(\eta_{200},\gamma)=(10,1.1)$, $(10,1.3)$, $(8.7,5/3)$ and
  $(6.8,5/3)$. Notice that for the same increase in $E_{\rm gas}$,
  increasing $\gamma$ has a greater effect on densities at small radii
  than reducing $\eta_{200}$. Finally, the dot-dashed curve
  uses $\gamma=1.2$ and $\eta_{200}=28$, and is representative of default
  profiles obtained in Models C and D.}
\label{figrhorange}
\end{figure}

\begin{figure}
\centerline{\psfig{figure=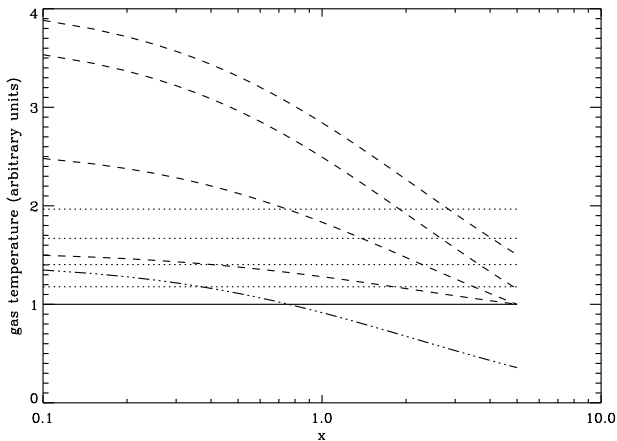,width=\figwidth}}
\caption[Examples of model temperature profiles]
  {As Fig.~\ref{figrhorange}, but showing temperature
  profiles (note that the temperature scale is linear).
  Temperatures have been normalized so that the
  solid curve has a temperature of unity. Increasing $\gamma$ leads to
  steeper temperature gradients without changing the temperature at
  the virial radius ($x=c$). In contrast, reducing $\eta_{200}$
  increases the temperature uniformly. }
\label{figtrange}
\end{figure}

\begin{figure}
\centerline{\psfig{figure=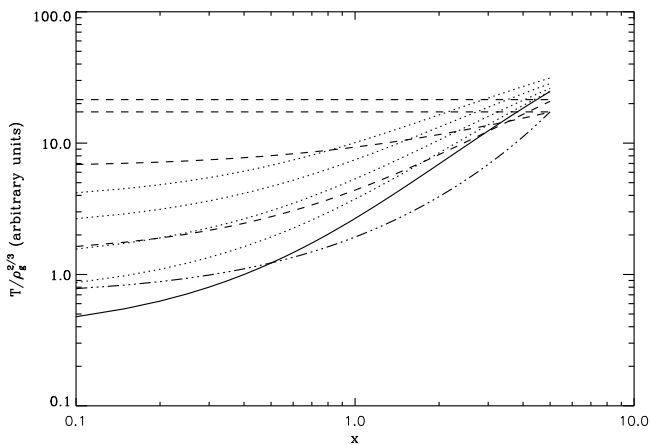,width=\figwidth}}
\caption[Examples of model entropy profiles]
  {As Fig.~\ref{figrhorange}, but showing `entropy'
  profiles, given by the expression $T/\rho_{\rm g}^{2/3}$. 
  Most of the profiles are quite
  steep, but increasing $\gamma$ (dashed curves) results in much
  flatter entropy profiles. Isentropic profiles are obtained when
  $\gamma=5/3$.}
\label{figentrange}
\end{figure}

In Figs.~\ref{figrhorange}, \ref{figtrange} and \ref{figentrange} we
display the density, temperature and entropy profiles of a selection
of gas halos covering a range of $\eta_{200}$ and $\gamma$ values. All
other parameters, in particular the total gas mass and NFW potential
well, have been kept constant. In each figure, the series of dotted
curves and dashed curves represent two different ways of heating the
gas halo represented by the solid curve.  In each series, the value of
$E_{\rm gas}$ was required to increase at regular intervals from that
of the solid curve up to a value of zero. Hence the gas halo with the
most energy in each series is only marginally bound. By comparing the
two series it is evident that profiles with the same total energy can
differ significantly.

\subsection{Profile selection: the heating models} \label{picksec}

\subsubsection{The default profile}

The first step is to determine the default profile.
In a 2-parameter family of gas profiles, a profile can be specified by
the value of $E_{\rm gas}$ and one further constraint. We shall
consider two different constraints for selecting default profiles:
$\gamma=1$ or $\gamma=1.2$, depending on the heating model.  The
former yields isothermal gas halos in the absence of heating, and is
motivated by its simplicity. The latter is motivated by the
temperature profiles of X-ray clusters measured by Markevitch et al.\ 
(1998\nocite{mfsv98}), who approximated their results
with a polytropic index of 1.2--1.3. For each constraint, we need to
calibrate the parameter $K$ used in equation~\ref{energy}.

We calibrate $K$ by matching the model clusters obtained with $E_{\rm
  excess}=0$ to the largest observed clusters. We do not attempt to
estimate $K$ theoretically, as it is our opinion that $E_{\rm gas}$
depends on how the collapse occurred in detail. For example, how the
gas collapsed relative to the dark matter affects how much energy was
transferred between the two components. [However, we do assume that
such processes result in the scaling law expressed in (\ref{energy}).]

To calibrate $K$ for the case of $\gamma=1$, we use the results of
Ettori \& Fabian (1998\nocite{ef98}), who fitted the surface
brightness profiles of 36 X-ray clusters with
$L_{\rm X}\ga 10^{45}$ \ergps.  When fitting to avoid
any cooling flow region, they obtain a mean value of $\eta=10.29$
with an rms scatter of 1.55.  (Since the temperature is constant,
$\eta$ and $\eta_{200}$ are the same.) In order to match this
we set $K=1.2$, which gives a mean value of $\eta=10.5$ in the
corresponding model clusters. However, the scatter of $\eta$ in our
model is only $\sim 0.5$.
If we now set the gas fraction of clusters equal to 0.17 (the mean value
obtained by \citeNP{evrar97} and \citeNP{ef98}, assuming $h=0.5$),
we find that the model clusters naturally follow the
observed $L_{\rm X}-T$ relation for
clusters more luminous than $2\times 10^{45}$ \ergps\
\cite{af98x}. (We refer to bolometric luminosities
throughout.) Note that this fit is possible because the largest observed
clusters roughly follow the self-similar relation $L_{\rm X}\propto
T^2$, instead of the steeper relation obeyed by smaller
clusters.

Turning to the case of $\gamma=1.2$, we note that compared to the
isothermal profiles these are
almost always poorer fits to the surface brightness profiles of real
clusters \cite{ef98}. Hence for this case we
calibrate $K$ by simply matching the $L_{\rm
  X}-T$ relation measured by \citeN{af98x}. As above, we set the
gas fraction of all clusters equal to 0.17.  We find that
$K=1.5$ results in an $L_{\rm X}-T$ distribution that best fits the
data. The resulting clusters have $\eta_{200}\approx 28$.  An example
of such a profile is shown in Figs.~\ref{figrhorange} to
\ref{figentrange} as dot-dashed curves, for comparison with the solid
curves ($\gamma=1$ and $\eta_{200}=10$).  Notice that although the two
density profiles have different shapes, they roughly follow each
other and intersect at two points. [The higher value of $\eta_{200}=28$
merely implies that the temperature at $r_{200}$ is lower by a factor
of 2.8 compared to the $\eta_{200}=10$ case.]

Since $\gamma$ is fixed for both types of default profile, it is not
hard to show that $\eta_{200}$ is a function of the NFW concentration
$c$ only. We find that it is only a weakly increasing function of $c$
in both cases. Since the model clusters have a
relatively small scatter in $c$ and $\eta_{200}$, they are close to
self-similar when heating is absent.

\subsubsection{The heated profile}

\begin{figure}
\centerline{\psfig{figure=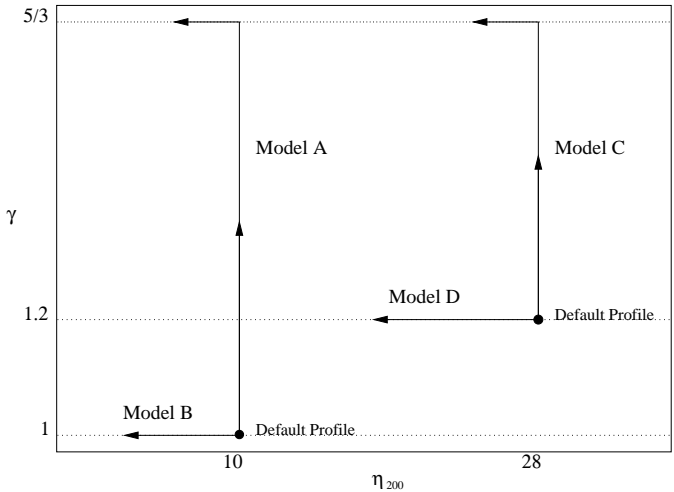,width=\figwidth}}
\caption[Schematic diagram of how gas profiles are selected in
each heating model] {A schematic diagram of how gas profiles are
  selected in each heating model. The first step is to find the
  default profile: depending on the heating model it has either
  $\gamma=1$ or $\gamma=1.2$; $\eta_{200}$ is then determined by
  requiring that the total specific energy, $E_{\rm gas}$, satisfies
  (\ref{energy}). (Note that the parameter $K=1.2$ for Models A and B,
  and $K=1.5$ for Models C and D.) The filled circles give only the
  approximate positions of default profiles, since $\eta_{200}$
  depends on the NFW concentration $c$. The heated profile is found in
  the second step: any excess specific energy increases $E_{\rm gas}$
  accordingly, and may increase the temperature uniformly (Models B
  and D), or increase the temperature preferentially towards the
  centre (Models A and C; these are modified to accommodate
  the upper limit of $\gamma=5/3$ when heating is very strong).}
\label{figmodel}
\end{figure}

When excess energy is present, the default profile is modified to give
the heated profile. We model this in two ways: by decreasing
$\eta_{200}$ while keeping $\gamma$ constant, or by increasing
$\gamma$ while keeping $\eta_{200}$ constant. The former has the
effect of increasing the temperature at all radii by the same amount
(to see this, multiply equation~\ref{polyt} by $T_{200}$ and note that
$\eta_{200} T_{200}$ remains constant).  The latter steepens the
temperature gradient while ensuring that the temperature at $r_{200}$
stays constant, so that heating is concentrated towards the centre.

Since there are two types of default profile, we have four heating
models in total. These are summarized in Fig.~\ref{figmodel}. Models A
and B have default profiles with $\gamma=1$ and Models C and D have
default profiles with $\gamma=1.2$. Heating increases $\gamma$ in
Models A and C, but reduces $\eta_{200}$ in Models B and D. 

There are a few loose ends to tie up. If the excess energy is so high
that $E_{\rm gas}>0$, then the gas is not bound and it does not form a
halo. However, for Models A and C, the gas halo may still be bound
when $\gamma$ has increased to 5/3.  Therefore, to increase $E_{\rm
  gas}$ further we reduce $\eta_{200}$ instead, as shown in
Fig.~\ref{figmodel}.

\section{THE EXCESS ENERGIES REQUIRED IN X-RAY CLUSTERS} \label{clustersec}

In this section, we present the cluster results obtained with each of
the four heating models. Since we are concerned solely with clusters
here, the parameter $\tau_0$ and the star formation model play almost
no part in the results. (No cold collapse occurs in the model clusters
for all reasonable values of $\tau_0$.)

The simulations are `low resolution' in the sense that
they only use the top 10 levels of the collapse tree
(section~\ref{treesec}). Hence, the smallest regions have masses of
$1.5\times 10^{13}$\Ms. Each simulation used a total of 10000
realisations of the merger tree.  We set the gas fraction
of every cluster equal to 0.17 \cite{evrar97,ef98} for definiteness.
The formulae used to calculate bolometric luminosity $L_{\rm X}$,
emission-weighted temperature $T$ and the instantaneous mass
deposition rate, $\dot{M}$, are given in Appendix A.  All
quantities were evaluated at $z=0$.  

One simulation was performed for each heating model, and in each case
all of the clusters were given a constant excess specific energy.  For
each heating model, we found the excess specific energy that best fit
the data by matching to the $L_{\rm X}-T$ relation of David et al.\ 
(1993\nocite{dsjfv93}) in the first instance.

\begin{table}
\caption[Best fitting values of excess energy for all four heating
models, obtained by matching model clusters to the observed $L_{\rm
  X}-T$ relation]
  {Best fitting values of excess energy for each heating model,
  obtained by matching to the $L_{\rm X}-T$ relation measured by David et
  al.\ (1993). Excess energy per particle is
  calculated as $(\mu m_{\rm H}\, \Delta E_{\rm gas})$.} 
  \nocite{dsjfv93}    \label{tabexcess}
\begin{center}
\vspace{0.5cm}
\begin{tabular}{cc}
  Heating Model & Excess Energy (keV/particle) \\
  \hline
  A & 1.8 \\
  B & 2.8 \\
  C & 2.2 \\
  D & 3.0 \\
  \hline
\end{tabular}
\end{center}
\end{table}

\begin{figure}
\centerline{\psfig{figure=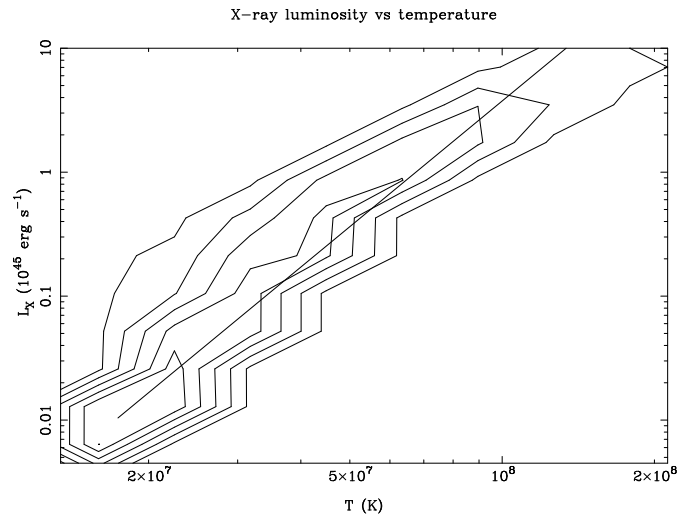,width=\figwidth,angle=270}}
\caption[The cluster $L_{\rm X}-T$ distribution from
Model A, with heating included] {Contour plot of the cluster X-ray
  luminosity-temperature distribution obtained from Model A, with
  heating included at the level given in Table~\ref{tabexcess}. The
  contours are spaced at equal logarithmic intervals. The long
  straight line is the best fit (for bolometric luminosities) taken
  from David et al.\ (1993). The extent of the line corresponds
  roughly to the extent of the data.} \nocite{dsjfv93}
\label{figlta}
\end{figure}

\begin{figure}
\centerline{\psfig{figure=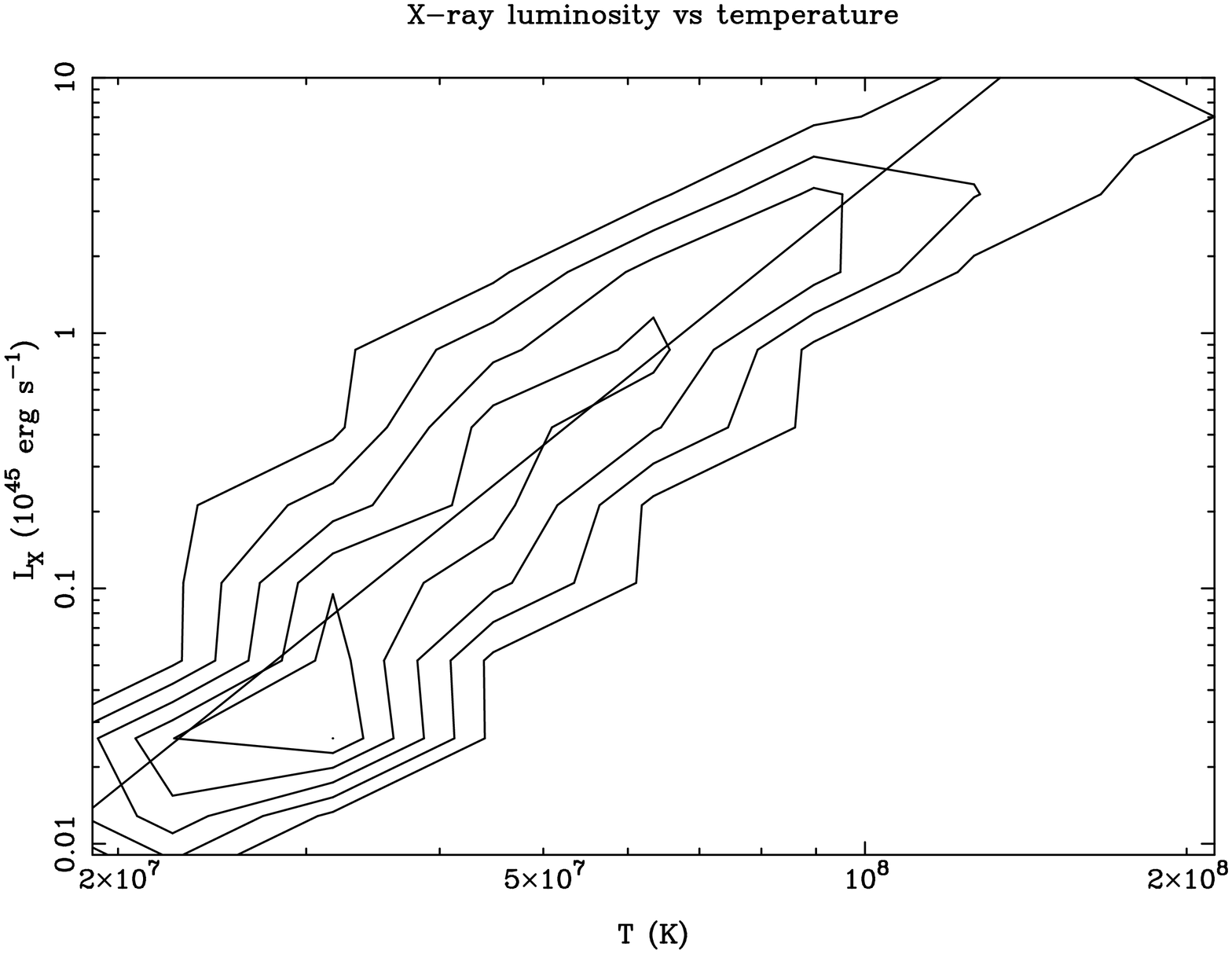,width=\figwidth,angle=270}}
\caption{Same as Fig.~\ref{figlta} but using Model B.}
\label{figltb}
\end{figure}

\begin{figure}
\centerline{\psfig{figure=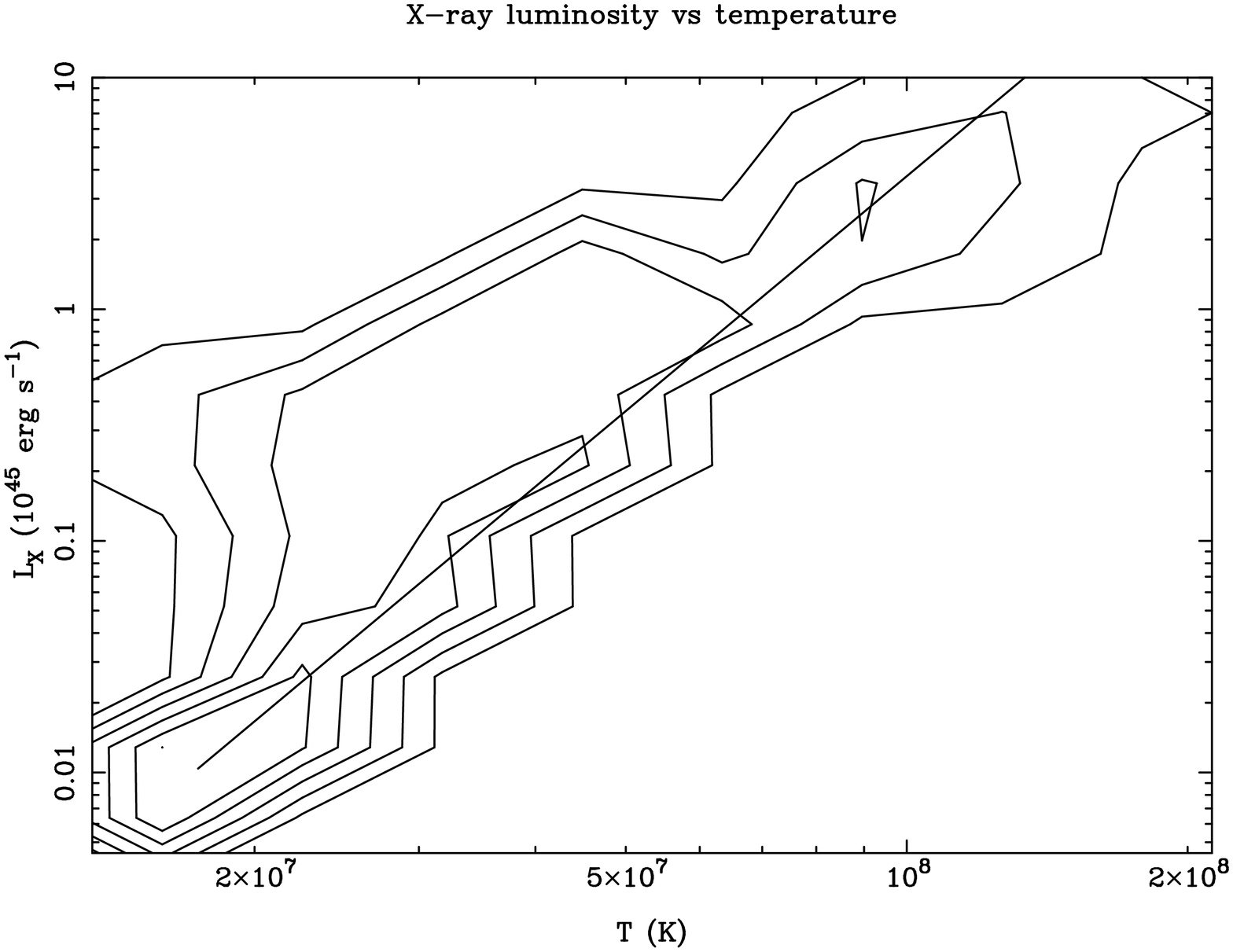,width=\figwidth,angle=270}}
\caption{Same as Fig.~\ref{figlta} but using Model C.}
\label{figltc}
\end{figure}

\begin{figure}
\centerline{\psfig{figure=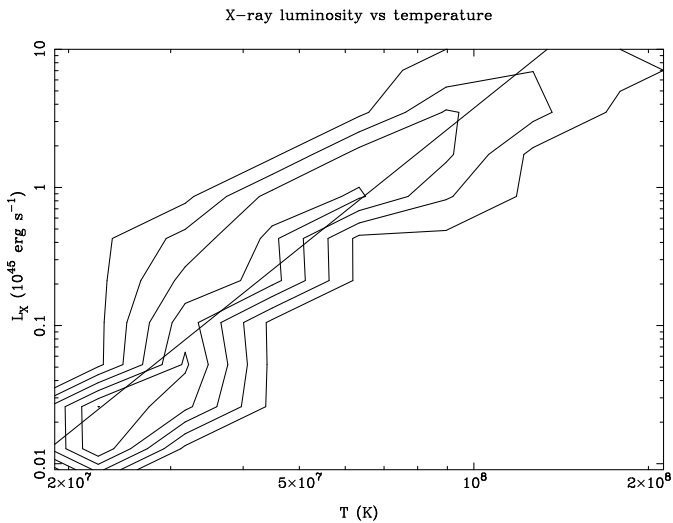,width=\figwidth,angle=270}}
\caption{Same as Fig.~\ref{figlta} but using Model D.}
\label{figltd}
\end{figure}

The best-fitting excess energy for each heating model is given in
Table~\ref{tabexcess}.  The resulting $L_{\rm X}-T$ distributions are
displayed in Figs~\ref{figlta} to \ref{figltd}.  The slopes of the
distributions given by Models B and D are slightly steeper than the
observed slope.  This suggests that we need to relax our assumption of
a constant $E_{\rm excess}$ for all clusters. It is also evident that
the $L_{\rm X}-T$ distributions flatten slightly at high temperatures,
tending to $L_{\rm X}\propto T^2$, in agreement with the largest
observed clusters (\citeNP{af98x}). 

Recall that we calibrated the
largest clusters to match the $L_{\rm X}-T$ relation of Allen \&
Fabian when heating is absent. The results thus confirm that the
largest clusters are least affected by the excess energy (see also
Fig.~1 in WFN98).  However,
the hottest clusters shown are in fact about a factor of 1/3 less
luminous than before heating. We do not attempt to correct for this
relatively small discrepancy.
It is possible that in reality $E_{\rm excess}$ would be more
diluted, \ie smaller than we have assumed, in the largest
clusters.
   
As expected, Models A and C require less heating than the other
models, because they concentrate heating towards the centre of
clusters, where most of the luminosity comes from. In addition, Models
C and D require slightly more excess energy than Models A and B, respectively.
Nevertheless, the highest excess energy in Table~\ref{tabexcess} is only
about 50 per cent more than the lowest, over a set of very
different heating models.

\begin{figure}
\centerline{\psfig{figure=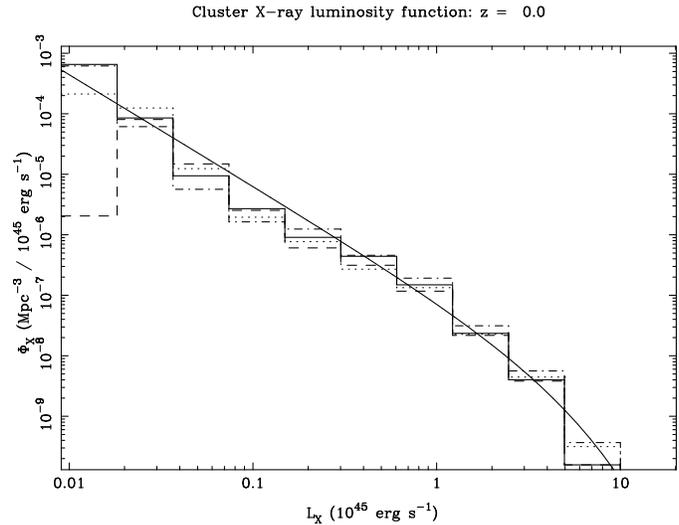,width=\figwidth,angle=270}}
\caption[The X-ray luminosity functions given by all four heating models]
  {The X-ray luminosity functions given by all four heating
  models. The model results are plotted as follows: Model A: solid line,
  Model B: dashed line, Model C: dot-dashed line, Model D: dotted
  line.  The curve is the best-fitting Schechter function for
  the {\em ROSAT} Brightest Cluster Sample bolometric luminosity
  function (Ebeling et al.\ 1997).}  \nocite{eefac97}
\label{figlxfunc}
\end{figure}

\begin{figure}
\centerline{\psfig{figure=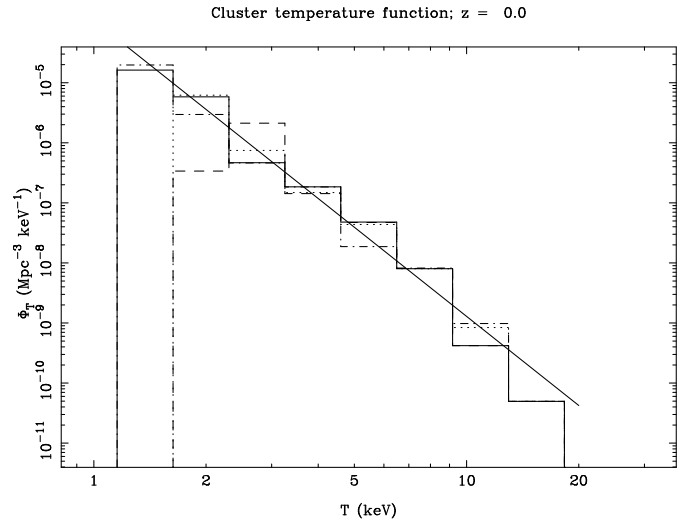,width=\figwidth,angle=270}}
\caption[The X-ray temperature functions given by all four heating models]
  {The X-ray temperature functions given by all four heating
  models, plotted with the same line styles as in
  Fig.~\ref{figlxfunc}. The straight line is the power law fit
  obtained by Edge et al.\ (1990).} \nocite{esfa90}
\label{figtfunc}
\end{figure}

\begin{figure}
\centerline{\psfig{figure=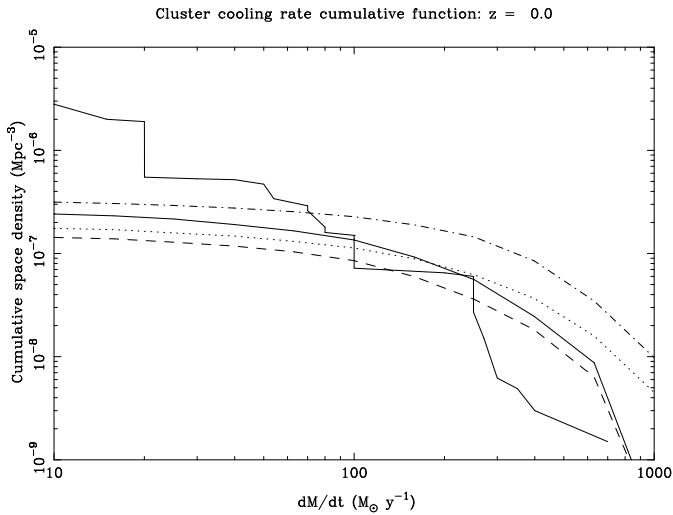,width=\figwidth,angle=270}}
\caption[The mass deposition rate functions given by all four heating models]
  {The mass deposition rate functions (plotted cumulatively) for
  all four heating models, plotted with the same line styles as in
  Fig.~\ref{figlxfunc}. The jagged line is the same
  function taken from Peres et al.\ (1997), modified by using a
  cluster age of 6 Gyr. 
  } \nocite{peres98}
\label{figmdfunc}
\end{figure}

We display the X-ray luminosity function, temperature function and
mass deposition rate ($\dot{M}$) function from the same simulations in
Figs.~\ref{figlxfunc}, \ref{figtfunc} and \ref{figmdfunc},
respectively. In each plot we have used a different line for each
heating model. Superimposed on each plot is the observed data, as
described in the captions. The same remarks regarding the simulated
and observed $\dot{M}$ functions made in the previous section apply
here. (However, the exclusion of clusters cooler than $2\times 10^7$K
has practically no effect on the $\dot{M}$ functions simulated here,
for most gas halos below this temperature have been unbound.)

The luminosity and temperature functions obtained with all four
heating models give good fits to the data. However, the model
$\dot{M}$ functions give relatively poor fits.

Models C and D give particularly poor fits where $\dot{M}>100$\Ms\ 
y$^{-1}$. This is because the mass deposition rate of large clusters
are too high in these models. This can be attributed to the flatter
cores of their gas density profiles. The poor performance of Models C
and D support the result that the $\gamma=1.2$ gas profiles are
relatively poor fits to the surface brightness profiles of large
clusters compared to the $\gamma=1$ profiles \cite{ef98}.

\begin{figure}
\centerline{\psfig{figure=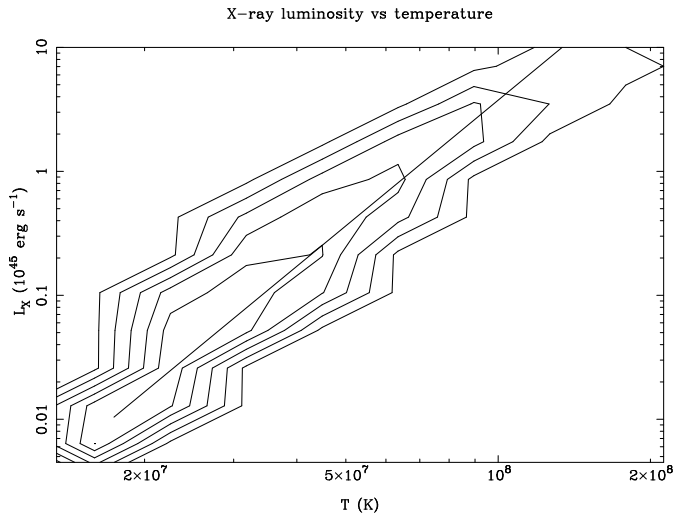,width=\figwidth,angle=270}}
\caption[Improved $L_{\rm X}-T$ distribution (Model B) using excess
energies which increase with halo mass, as given in Table~\ref{tabx}]
{As Fig.~\ref{figltb}, but using excess energies which increase with
  halo mass, as given in Table~\ref{tabx}. Model B was used.
  Previously unbound groups now appear at temperatures below 2 keV.}
\label{figltbnew}
\end{figure}

\begin{figure}
\centerline{\psfig{figure=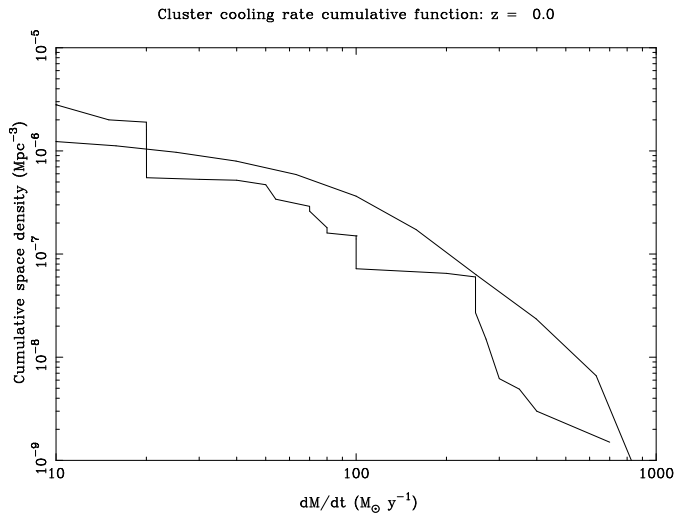,width=\figwidth,angle=270}}
\caption[The mass deposition rate function from the same simulation as
Fig.~\ref{figltbnew}]
  {As Fig.~\ref{figmdfunc} but for Model B only, using increasing
  excess energies with halo mass as given in Table~\ref{tabx}.
  The number of small cooling flows has increased, improving the fit
  to the data.}
\label{figmdbnew}
\end{figure}

Models A and B show a deficit of clusters with small
cooling flows ($\dot{M}=10$--100\Ms\ y$^{-1}$). 
The main reason for the deficit is because the excess energies are now
too high for the smallest clusters. We have repeated the simulation
for Model B using lower excess energies in clusters less massive than
$246\times 10^{12}$ \Ms.  The excess energies are given in
Table~\ref{tabx}; they increase steadily with mass up to $246\times
10^{12}$ \Ms. The resulting $L_{\rm X}-T$ distribution and $\dot{M}$
function are shown in Figs.~\ref{figltbnew} and \ref{figmdbnew},
respectively. Both show a better match to the data than before. The
new $\dot{M}$ function has an increased number of small cooling flows,
and the new $L_{\rm X}-T$ distribution reaches to lower temperatures
(due to the reappearance of $\sim 2$ keV clusters, which were previously
unbound).

\begin{table} 
\caption{Table of excess energies used with Model B to improve the
  mass deposition rate function, which is shown in
  Fig.~\ref{figmdbnew}.\label{tabx}}

\begin{center}
\vspace{0.5cm}
\begin{tabular}{cc}
  Halo Mass ($10^{12}$\Ms) & Excess Energy (keV/particle) \\
  \hline
  $\geq 246$ & 2.8 \\
  123        & 2.3 \\
  61         & 1.9 \\
  35         & 1.5 \\
  15         & 1.1 \\
  \hline
\end{tabular}
\end{center}
\end{table}

If it is true that $E_{\rm excess}$ increases with cluster mass, then
this may be hard to reconcile with heating by supernovae, because we
then expect $E_{\rm excess}$ to become more diluted with increasing
halo mass (see section~\ref{galsec}). In this case, a significant
amount of energy injection would have to occur in clusters themselves
(possibly by AGN). However, we note that this result is somewhat
model-dependent, for it is possible to avoid it by combining different
heating models.  If large clusters are heated preferentially towards
the centre (as in Model A) but small clusters are heated more
uniformly (as in Model B), then it is possible that an excess energy
of roughly 1.8 keV/particle across all clusters could satisfy all the
data (see Table~\ref{tabexcess} and \ref{tabx}). Such a scenario may
result from a characteristic scale in the spatial distribution of the
heat source (supernovae or AGN). Alternatively, a strong wind may
distribute its energy more efficiently through a small
(proto-)cluster, because the cluster is closer to being unbound, \ie
it is more disturbed.

\subsection{Using all available gas profiles} \label{apsec}
By using all the available gas profiles in the 2-parameter family (\ie
independently of any heating model), we have also found the minimum
excess energy required to put a fiducial cluster on the observed
$L_{\rm X}-T$ relation.

\begin{figure}
\centerline{\psfig{figure=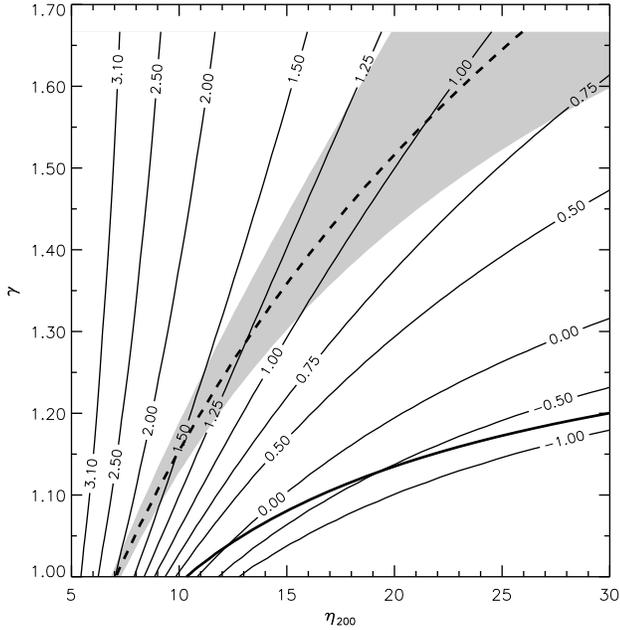,width=\figwidth}}
\caption[Contour plot of excess energy in parameter space, for a
fiducial cluster of mass $1.23\times 10^{14}$\Ms, showing also the
parameters which satisfy the observed $L_{\rm X}-T$ relation] {Contour
  plot in parameter space for a fiducial cluster of mass $1.23\times
  10^{14}$ \Ms, collapsing at $z=0$ with a gas fraction of 0.17. The
  dashed curve gives the parameters of gas halos that lie on the
  best-fitting $L_{\rm X}-T$ relation obtained by David et al.\ 
  (1993), and gas halos in the shaded region lie within the region of
  uncertainty of this relation. The thin contours are labelled by
  excess energy (keV/particle), measured relative to an isothermal
  default profile (as in Models A and B).  The profile that requires
  the least excess energy to match the $L_{\rm X}-T$ relation is given
  by $\gamma=5/3$ and $\eta_{200}=26$.  It has an excess energy of
  0.95 keV/particle.  The thick solid line roughly sweeps out the
  positions of other possible default profiles (see text).}
\label{figetagamtab}
\end{figure}

We considered the specific case of a halo of mass $1.23\times 10^{14}$
\Ms, which virializes at $z=0$ with a gas fraction of 0.17.  Such a
cluster has a temperature of around 2 keV, depending on the amount of
heating.  To obtain the NFW profile we assumed the same cosmology as
before. The problem was structured as follows. We first found the
locus of points in $(\eta_{200},\gamma)$-space which put the cluster
on the observed $L_{\rm X}-T$ relation.  From these points we then
found the one which had the least excess energy.  However, the gas
profile specified by $(\eta_{200},\gamma)$ only tells us the value of
$E_{\rm gas}$---to compute $E_{\rm excess}$ we also need the `default'
value of $E_{\rm gas}$, \ie when heating is absent. In what follows,
we assume that the default value of $E_{\rm gas}$ is given by
equation~\ref{energy} with $K=1.2$ (as in Models A and B). 

Fig.~\ref{figetagamtab} shows contours of excess energy in parameter
space, labelled in keV/particle. The gas halo becomes unbound for
excess energies above 3.1 keV/particle.  The dashed curve gives the
parameters of gas halos that lie on the best-fitting power law to the
observed $L_{\rm X}-T$ distribution \cite{dsjfv93}. The shaded area
contains gas halos that lie within the $1\sigma$ region of uncertainty
for this best-fit. 
Note that it represents the uncertainty in the {\em mean} properties
of X-ray clusters, and should not be confused with the dispersion in
the $L_{\rm X}-T$ relation.  From the plot, the gas profile with
$\gamma=5/3$, $\eta_{200}=26$ requires the least excess energy to
match the best-fit relation. It has an excess energy of 0.95
keV/particle. If the shaded region is taken into account, the minimum
excess energy is roughly 0.7 keV/particle.  It should not be
surprising that the above profile is marginally stable to convection.
We `save energy' by concentrating the heating where it makes the most
difference, \ie near the centre, but convection limits the extent to
which we can do this.  The gas halo that requires the least heating is
therefore the one with the isentropic atmosphere.  This suggests that
the $\gamma=5/3$ profile probably requires the least heating among all
possible gas profiles.

\begin{figure}
\centerline{\psfig{figure=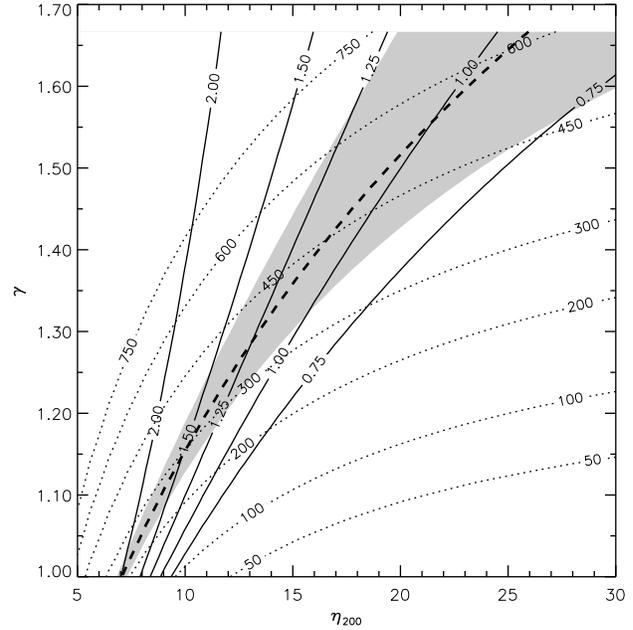,width=\figwidth}}
\caption[Similar to Fig.~\ref{figetagamtab}, showing also contours
of entropy at a radius of $0.1 r_{200}$]
{Similar to Fig.~\ref{figetagamtab}, showing also contours
of entropy at a radius of $0.1 r_{200}$ (dotted lines). The contour
labels give $s=T/n_e^{2/3}$ in units of keV~cm$^2$.}
\label{figetagamtab2}
\end{figure}

A similar plot displayed in Fig.~\ref{figetagamtab2} shows contours of
entropy (given by $s=T/n_e^{2/3}$) at a radius of $0.1 r_{200}$. The
entropy varies significantly along the dashed line, from around 200
keV~cm$^2$ to 600 keV~cm$^2$. The plot shows that the energy
requirements are reduced if heating raises the entropy as much as
possible.  (We discuss this further in section~\ref{ponmansec}.)  The model
entropies may be compared to the results of \citeN{pcn99}, who
measured the entropies of groups and clusters at this radius in order
to avoid possible cooling flows. However, these authors used
emission-weighted temperatures to compute the entropy, whereas we have
used radial-resolved temperatures. When this is accounted for, the
above range of entropies are all consistent with the data.

So far we have assumed that in the absence of heating $E_{\rm gas}= -3.1$
keV/particle, as given by equation~\ref{energy} using $K=1.2$. If we
use $K=1.5$ instead (as in Models C and D), then the default value of
$E_{\rm gas}$ becomes $1.5/1.2\times (-3.1)\approx -3.9$ keV/particle. As
this is lower than before, all excess energies are
{\em increased} by 0.8 keV/particle.
We can generalize further by considering what parameters the cluster
would need in order to lie on the self-similar relation $L_{\rm
  X}\propto T^2$ normalized to the largest observed clusters \cite{af98x}.
The gas profiles which satisfy this relation are given by the thick
solid line
in Fig.~\ref{figetagamtab}. As expected, it passes close to the points
$(\eta_{200},\gamma)=(10,1)$ and (28,1.2), where the default
profiles of our heating models are found. 
Thus the thick line roughly sweeps out the locations of possible
default profiles. By assuming a $\gamma=1$ default profile in the
above analysis, we obtained the highest default value of
$E_{\rm gas}$, and therefore the lowest possible {\em excess} energies.

\section{THE EFFECT OF SUPERNOVA HEATING} \label{galsec}

In this section, we investigate the amount of excess energy obtainable
from supernova heating.
Complete simulations with 20 levels of collapse hierarchy
were performed with Models A and B. Each simulation used 100
realisations of the merger tree. Below, we begin by setting the
parameters of the galaxy formation model.

\subsection{Setting the model parameters} \label{fixsec}

There are three parameters that remain to be set. They
are the critical ratio of cooling time to free-fall time, $\tau_0$,
the efficiency of supernova feedback, \esn, and the boost in the rate
of supernovae. As mentioned in section~\ref{sfsec}, we assume that
supernova rates are boosted by a factor of 5 for this
work. Intuitively, this should increase the amount of supernova
heating; however, we shall demonstrate that the resulting excess
energies are quite insensitive to this parameter. All three
parameters are kept constant in each simulation.

We assume an initial gas fraction of 0.27 (section~\ref{introsec}).
Unless stated otherwise, the resulting X-ray clusters have a mean gas
fraction of 0.17 and a scatter of about 0.01, in agreement with the
gas fraction used in the previous section.

\subsubsection{Setting \esn}
The feedback parameter \esn\ controls the amount of star formation,
which can be characterized by the fraction of gas turned into stars by
the present day. Using the Coma cluster as a large sample of baryons,
the mass ratio of hot gas to stars inside a radius of $1.5 h^{-1}$ Mpc
is about 15, assuming $h=0.5$ \cite{wnef93}. In order to match this, we
set \esn=0.3 for Model A and \esn=0.25 for Model B. We find that the
required value of \esn\ is almost independent of the value of
$\tau_0$, unless $\tau_0$ takes an `extreme' value ($\sim 10$ times
greater or smaller than 1).
In fact, a much larger fraction of baryons is converted into BDM than into
stars (as can be seen from the primordial and cluster gas fractions).
Most of the BDM is formed in the halos of massive galaxies and small
groups.

\subsubsection{Setting $\tau_0$} 
The parameter $\tau_0$ controls the transition from cold to hot
collapse. From its definition, we know that $\tau_0\sim 1$.  However,
we consider a range of values: $\tau_0=1$, 0.4, and an extremely low
value of 0.1, to illustrate its effect on the resulting excess
energies.  Table~\ref{tabte} lists the three sets of parameters used
in the simulations.

\begin{table}
  \caption[The values of the parameters \esn\ and $\tau_0$ used with
  Models A and B]
  {The values of \esn\ and $\tau_0$ used with Models A and B.} \label{tabte}
  \vspace{0.5cm}
  \begin{center}
    \begin{tabular}{lll}
            & \esn & $\tau_0$ \\ 
      \hline
      Model A & 0.3  & 1.0      \\
      Model B & 0.25 & 0.4     \\ 
      Model B & 0.15 & 0.1     \\
      \hline
    \end{tabular}
  \end{center}
\end{table}

\subsection{The excess energies from supernova heating} \label{galsec1}

\begin{figure}
\centerline{\psfig{figure=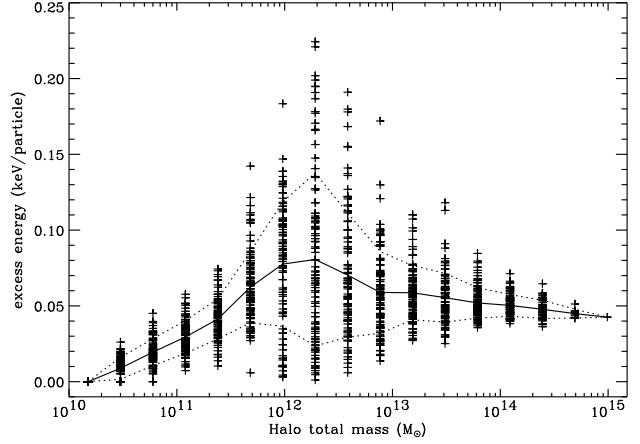,width=\figwidth}}
\caption[Scatter plot of excess energy from supernova heating as a
function of halo mass, using Model A] {Scatter plot of excess energy
  vs.\ halo mass, using Model A. Each mass bin contains a maximum of 100
  points. Halos were selected randomly from the simulation regardless
  of redshift. The solid line gives the mean value of each mass bin,
  and the dotted lines are plotted one standard deviation from the
  mean.}
\label{figexcess} 
\end{figure}

\begin{figure}
\centerline{\psfig{figure=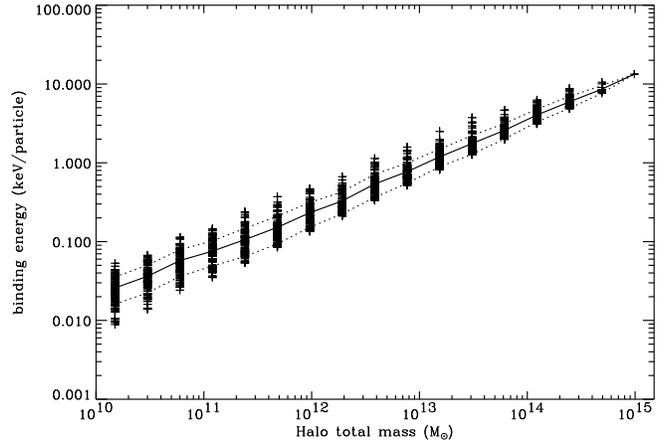,width=\figwidth}}
\caption[Scatter plot of binding energy vs.\ halo mass, using Model A]
  {As Fig.~\ref{figexcess}, but showing the magnitude of the binding
  energy vs.\ halo mass. Note that the energy is now plotted logarithmically.}
\label{figbe}
\end{figure}

\begin{figure}
\centerline{\psfig{figure=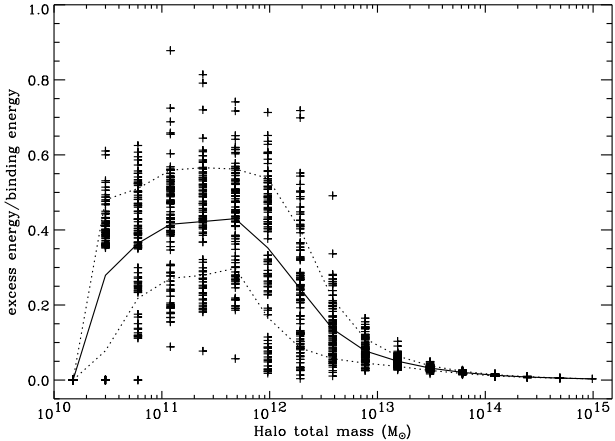,width=\figwidth}}
\caption[Scatter plot of the ratio of excess energy to binding energy
vs.\ halo mass, using Model A] {As Fig.~\ref{figexcess}, but showing
  the ratio of excess energy to binding energy. The dip in the mean
  for the lowest mass bins is caused by halos which have zero excess
  energy. This can only occur when a halo has no progenitors.
  Therefore, the dip is an artifact of the finite mass resolution.
  Notice that the finite points in the second lowest mass bin
  already have a similar distribution to higher mass bins.}
\label{figexoverbe}
\end{figure}

For Model A, a scatter plot of excess energies vs.\ halo mass is
displayed in Fig.~\ref{figexcess}, along with the mean and standard
deviation for each mass bin. All of the scatter plots in this section
were generated by randomly selecting up to 100 halos for each mass,
regardless of redshift (only the most massive halos have less than 100
points plotted, because they are so rare).  Up to a mass of $\sim
10^{12}$\Ms, the excess energies clearly increase with mass.  Above
$\sim 10^{12}$\Ms, star formation gives way to cooling flow behaviour,
so that the mean excess energy changes little.  However, the scatter
reduces significantly due to an averaging effect. A gradual decrease
in excess energy can be detected in the most massive halos, due to
dilution by the accretion of primordial gas.

The ratio of excess energy to binding energy gives a measure of the
excess energy's ability to change the gas distribution.  Recall that
we define the binding energy to be equal to $|E_{\rm gas}|$ in the
absence of heating.  Fig.~\ref{figbe} shows a corresponding plot of
binding energy for the same simulation as above.  The ratio of excess
energy to binding energy is displayed in Fig.~\ref{figexoverbe}. It
has a strict upper limit of 1, above which gas halos are not bound.
The distributions of points in mass bins below $\sim 10^{12}$\Ms\ are
very similar and lie roughly in the range 0.2--0.6. The lowest mass
bins are an exception because some of their halos have no excess
energy at all; this causes the mean to dip for the lowest bins. As
explained in the caption, this is purely an artifact of the finite
mass resolution.

The approximately scale-invariant behaviour below $\sim 10^{12}$\Ms\ 
can be understood as follows.  Below a certain halo mass, almost all
of the galaxies produce sufficient supernova feedback to eject their
gas.  In addition, the gas is always ejected with excess energy equal
to the binding energy of the host halo (for the model assumes that the
gas is ejected at the escape velocity). As a result, for halos in, or
slightly above, the said mass range, the ratio of excess energy to
binding energy simply reflects the ratio of the binding energies of
its progenitors to itself (ignoring the dilution of excess energy by
primordial gas for simplicity). The similarity in the distribution of
points in each mass bin (below $\sim 10^{12}$\Ms) simply implies that
these ratios do not change much with mass. 

For halos $\ga 10^{12}$\Ms,
the ratio drops dramatically due to the cessation of star formation.
Above $10^{14}$\Ms---in the regime of X-ray clusters---the excess
energies have hardly any effect on the gas halos.

\begin{figure}
\centerline{\psfig{figure=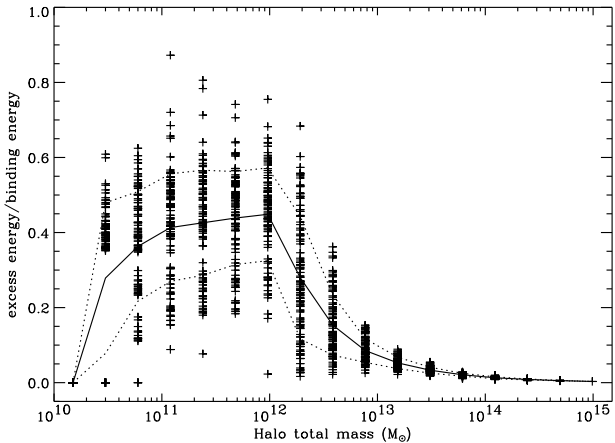,width=\figwidth}}
\caption[Scatter plot of the ratio of excess energy to binding energy,
using Model B with $\tau_0=0.4$]
  {The ratio of excess energy
  to binding energy obtained from Model B, using $\tau_0=0.4$ and
  \esn=0.25. Note the strong similarity with Fig.~\ref{figexoverbe}.
}
\label{figexoverbeb}
\end{figure}

\begin{figure}
\centerline{\psfig{figure=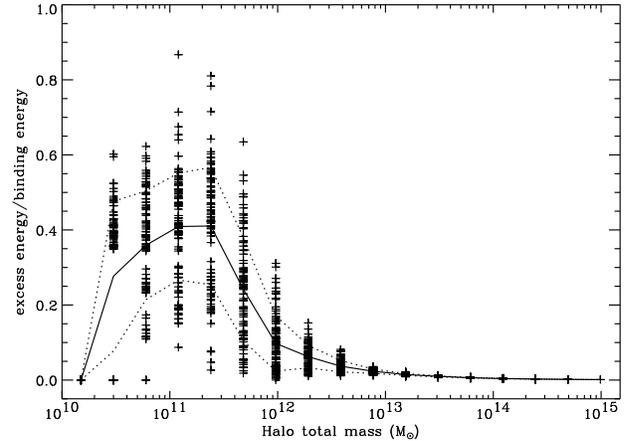,width=\figwidth}}
\caption[Scatter plot of the ratio of excess energy to binding energy,
using Model B with $\tau_0=0.1$]
  {The ratio of excess energy
  to binding energy obtained from Model B, using $\tau_0=0.1$ and
  \esn=0.15. The lower value of $\tau_0$ causes star formation to
  cease at lower masses, resulting in a notable difference from
  Fig.~\ref{figexoverbeb}.
  }
\label{figexoverbebb}
\end{figure}

Since the behaviour shown in Fig.~\ref{figexoverbe} is largely due to
the binding energies of halos, it should depend little on the heating
model.  Fig.~\ref{figexoverbeb} shows the corresponding plot for Model
B, with the parameters $\tau_0=0.4$ and \esn=0.25. As expected, it is
almost the same as for Model A.  However, a difference does occur if
$\tau_0$ is reduced further. For Fig.~\ref{figexoverbebb}, we used the
parameters $\tau_0=0.1$ and \esn=0.15 with Model B. In this case, star
formation is restricted to much smaller halos, so that the decline is
also shifted to a lower mass scale. (A side effect is that more gas is
lost in cooling flows, so that the gas fraction of clusters is 0.15
instead of 0.17).  This scenario is unlikely to occur in reality, not
only because we expect $\tau_0\sim 1$, but also because the
characteristic luminosity, $L_*$, of the luminosity function of
galaxies (fitted with a Schechter function) would be too small.

\subsection{More on heating clusters with supernovae}
\label{galclussec}

The excess energies obtained above are clearly too low to satisfy the
energy requirements of X-ray clusters (section~\ref{clustersec}).  The
relationship between excess energies and binding energies also
suggests that it would be difficult to increase the amount of heating
significantly in this model. 
Indeed, we find that the excess energies of clusters
are {\em not sensitive} to \esn, nor the supernovae rate per unit star
formation. For example the parameters \esn=0.1 and \esn=1.0, used
with Model A, give virtually identical excess energies in clusters to
those shown in Fig.~\ref{figexcess}---indeed, the rest of the plot is
hardly modified. If instead we remove the factor-of-5 boost in
supernova rates (implying a change in the IMF), the excess energies of
clusters are only reduced from around $0.05$ to $0.03$ keV/particle.

Expanding on the previous section, the excess energy of a cluster is
essentially determined by the binding energies of the most massive
progenitors in its merger tree (looking backwards in time along each
and every branch) to produce type II supernovae. Although these
progenitors might not be able to eject their atmospheres, they are
still likely to leave the gas with $E_{\rm excess}$ close to the binding
energy.  Furthermore, the extent to which $E_{\rm excess}$ is diluted by
primordial gas in the final cluster is also mainly a function of the
merger tree. The net result is that changing \esn\ or the IMF has
little effect on the excess energy of clusters. What they do affect is
the amount of gas converted into stars:
the more efficient the supernova feedback, the less stars are formed.

The above suggests that we can increase the excess energies of
clusters by increasing $\tau_0$. We find that by increasing $\tau_0$
from 1 to 3, the transition from star formation to cooling flow
behaviour is shifted to halos that are roughly 4 times more massive.
As a result, $E_{\rm excess}$ in clusters increases from around 0.05
to 0.12 keV/particle. This agrees very well with a simple scaling
argument: since binding energy scales roughly as $M_{\rm tot}^{2/3}$,
the 4-fold increase in the mass scale of the transition region implies
that $E_{\rm excess}$ should increase by a factor of $4^{2/3}$; this
is indeed the case, but the increase is clearly too small.

\subsubsection{The simulated iron abundances}

The clusters shown in Figs.~\ref{figexoverbe} and \ref{figexoverbeb}
have an iron abundance of about 0.08\Zs. Although this is lower than
the observed range of 0.2--0.3\Zs\ \cite{fmtex98}, we reiterate that
this is not the reason for their low excess energies.  Like the
gas-to-stellar mass ratio, the iron abundance can be controlled by the
parameter \esn. For example, reducing \esn\ by a factor of 3 increases
both the stellar mass and the iron abundance of clusters by about a
factor of 3.

The large number of type II supernovae per unit stellar mass required
to enrich cluster gas to the observed metallicities has been discussed
by other authors \cite{arbvv92,eav95,bm98}. It is possible that a
large fraction of the iron in cluster gas is due to type Ia
supernovae, which we have not included. \citeN{ns98} suggest that
between 30--90 per cent of the iron in X-ray clusters may be due to
type Ia supernovae. It is also possible that the observed
metallicities (which are emission-weighted) overestimate the average
metallicities of cluster gas, due to the existence of steep
metallicity gradients \cite{efmot97,af98y}.

\section{Limitations of the model} \label{ensec}

In our model, we make the approximation that the excess specific
energy of a gas halo is equal to the total energy injected over
the history of the gas. \ie
\begin{equation}\label{estexc}
  E_{\rm excess}\approx{1\over M_{\rm gas}}
         \int\!\!\int\Gamma \,\d V \d t,
\end{equation}
where $M_{\rm gas}$ is the mass of the gas halo and $\Gamma$ is the
net heating rate per unit volume.  In general, $\Gamma$ thus includes
heating by supernovae and active galactic nuclei, and accounts for the
energy lost through radiative cooling. We refer to $\Gamma$ simply as
the rate of non-gravitational heating. The volume integration is made
over all of the gas that eventually forms the gas halo, therefore the
volume itself is irregular and varies with time.

However, there are mechanisms other than $\Gamma$ that can affect the
final value of $E_{\rm excess}$ and hence warrant at least a mention.
In what follows, we shall consider a single halo and the evolution
leading up to its virialization. We use the term `proto-halo' to refer
to the contents of this halo at all times earlier than the
virialization time
(note that the proto-halo is not itself a halo, but it can contain
progenitor halos).

Briefly, the mechanisms are as follows.
\begin{enumerate}
\item If the evolution of the gas distribution (which otherwise
  traces the DM distribution fairly well) is modified significantly by
  non-gravitational processes, then there can be a
  `gravitational contribution' to $E_{\rm excess}$.
\item If the gas pressure outside the proto-halo is raised
  significantly due to heating, then the work it does on the proto-halo may
  need to be included.
\item In any progenitor halo that contains hot gas, work is done (by the
  gas remaining) on gas that cools out near the centre. This has the
  effect of reducing $E_{\rm excess}$.
\item Gas that is converted to stars and BDM is generally
  located in positions of minimum potential. Removal of this gas may
  therefore increase the mean energy of the gas that remains.
\end{enumerate}
The mechanisms have been listed in order of increasing sophistication
in the arguments required. We consider each of them below and
attempt to quantify their effects on $E_{\rm excess}$. We also
give a more formal definition of $E_{\rm excess}$ and discuss the
evolution of $E_{\rm gas}$ in some detail. For definiteness, we shall base
our discussion on the proto-halo of a cluster, but it can be
generalised to smaller halos.

Quite aside from the effects mentioned above, there remains the
possibility that when the excess energy is large, some of the gas
associated with a DM halo may extend beyond the virial radius. Also,
there is some uncertainty in the efficiency with which gas that is
ejected from a halo recollapses into larger halos. We assumed that
such effects are small in our model.

If the heating of proto-cluster gas is very uneven, \eg if the gas is
heated by the radio jets of AGN, then the main effect may be to unbind
part of the intracluster medium. In this case, smaller clusters would
have lower gas fractions than larger clusters. However, in order to
match the observed $L_{\rm X}-T$ relation, the excess energies would
still need to be very high. The X-ray luminosity of a 2 keV cluster is
an order of magnitude below the self-similar prediction (see \eg
Fig.~1 of WFN98).
Since $L_{\rm X}$ scales as the gas density squared,
we would need to unbind 2/3 of the gas to reduce $L_{\rm X}$ by an
order of magnitude (assuming that the shape of the gas density profile
remains unchanged).  The excess energy averaged over all of
the gas is then $\approx 2/3$ of the binding energy of the cluster.

\subsection{The `gravitational' contribution} \label{mech1}

We begin with a simplified scenario in which no gas is converted into
stars or BDM in the proto-halo. We generalize the definition of
$E_{\rm gas}$ (equation~\ref{egas}) to apply to the proto-halo at
any time, by including the kinetic energy of bulk motion:
\begin{equation} \label{egasg}
    E_{\rm gas} \equiv 
  \frac{1}{M_{\rm gas}}\int\rho_{\rm g}\left(\frac{3kT}{2\mu m_{\rm H}}
  + \half\vel^2 + \phi\right) \,\d V,
\end{equation}
where $\vel$ is the velocity of the gas and the volume of integration
is as explained above.  At early times, the proto-halo occupies a
roughly spherical region; it later condenses into sheets, filaments
and halos.  As a first approximation, the potential $\phi$ can
therefore be calculated from the mass distribution of the proto-halo,
ignoring all matter outside it. (Using a larger region to calculate
$\phi$ does not affect our argument, but this simplifies estimates
of $E_{\rm gas}(t)$.) We set $\phi=0$ at infinity.

\begin{figure}
  \centerline{\psfig{figure=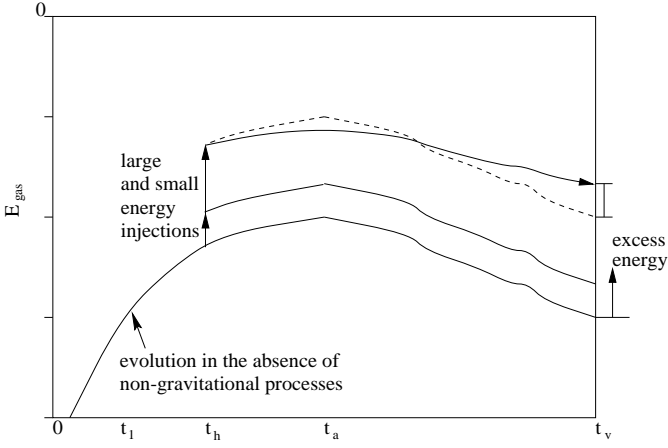,width=\figwidth}}
\caption[A schematic diagram of the evolution of $E_{\rm gas}$ with
time, with and without non-gravitational processes]
  {A schematic diagram of the evolution of $E_{\rm gas}$ with
  time. The times $t_h$, $t_a$ and $t_v$ give the time of energy
  injection, the turnaround time and the virialization time
  respectively. The lowest curve gives the evolution of $E_{\rm gas}$
  in the absence of non-gravitational heating or cooling. $E_{\rm
    excess}$ is defined as the deviation from this curve at $t_v$. The
  other two solid curves show the effects of injecting a small and large
  amount of energy. In the latter case, there is a `gravitational
  contribution' to the excess energy, given by the difference between
  the solid and dashed curves at $t_v$.}
\label{figensec}
\end{figure}

In Appendix B (equation~\ref{B1}) we show that $E_{\rm gas}$ obeys
\begin{equation} \label{egasevol}
{\d E_{\rm gas} \over\d t}
= {1\over M_{\rm gas}}
\int \left(\rho_{\rm g} \dpar{\phi}{t} + \Gamma \right) \, \d V,
\end{equation}
where we have assumed that the gas pressure at the boundary of the
proto-halo is negligible. This implies that the rate of change of
$E_{\rm gas}$ is given by the net rate of non-gravitational heating,
plus a weighted average of $\dpar{\phi}{t}$. Since $\phi$ is dominated
by the contribution from DM, we shall make the approximation
throughout that $\phi$ is unchanged by modifications in the gas
distribution.  This leads to the important observation that the gas
processes which drive $\Gamma$ do not have an immediate effect on the
other, gravitational term. (This would not be the case if, for
instance, that term included $\dpar{\rho_{\rm g}}{t}$ instead of
$\rho_{\rm g}$.) This allows us to consider the two terms on the rhs
separately.

In the absence of non-gravitational processes (implying $\Gamma=0$),
we expect $E_{\rm gas}$ to increase as the system expands,
and to decrease after the turnaround time, $t_a$. The final value
of $E_{\rm gas}$, at the virialization time $t_v$, is given by
equation~\ref{energy} in our model. A schematic diagram of this is
shown in Fig.~\ref{figensec}.  Equation~\ref{energy} itself simply
expresses how we expect $E_{\rm gas}$ to scale in the absence of
non-gravitational processes. 

The formal definition of $E_{\rm excess}$ is thus the difference
between the actual value of $E_{\rm gas}$ at $t_v$ and the value
obtained in the absence of non-gravitational processes.  Now suppose
the inclusion of non-gravitational heating does not modify the gas
distribution at all.  In this case, the gravitational term in
equation~\ref{egasevol} is not affected. $E_{\rm excess}$ is then
given by equation~\ref{estexc} exactly.  This is illustrated in
Fig.~\ref{figensec} by a single, small injection of energy at time
$t_h$. The subsequent evolution of $E_{\rm gas}$ is unchanged.

If the energy injected is large (comparable to $|E_{\rm gas}|$), then
it can make the gas distribution more extended in the potential well
of the proto-halo. This is likely to reduce the magnitude of the
gravitational term in equation~\ref{egasevol}, because more weight is
given to areas of smaller $|\phi|$, where $|\partial\phi/\partial t|$
is also likely to be smaller. The net change in $E_{\rm gas}$ between
energy injection and $t_v$ is therefore reduced. This is illustrated
by the solid curve resulting from the large injection of energy in
Fig.~\ref{figensec}.  Its deviation from the dashed curve (which would
describe $E_{\rm gas}$ if the gas distribution were not modified)
leads to an excess energy that is larger than the energy originally
injected.  We refer to the difference between the solid and dashed
curves at $t_v$ as the `gravitational contribution' to $E_{\rm
  excess}$.  In general, the gravitational contribution is given by
\begin{equation}
    {1\over M_{\rm gas}}
         \int\!\!\int \left( \rho_{\rm g} -
                      \rho_{\rm g,G}\right) \dpar{\phi}{t} 
         \,\d V \d t.
\end{equation}
Here and below, we use a subscript `G' to imply the same system
evolved without including non-gravitational processes.

The above argument suggests that provided $E_{\rm gas,G}(t_h)>E_{\rm
  gas,G}(t_v)$, the gravitational contribution associated with a large
injection of energy is likely to be positive, especially if the above
inequality is large. From Fig.~\ref{figensec} we can see that this is
no longer true if $t_h$ is earlier than some time $t_1$, given by
$E_{\rm gas,G}(t_1)=E_{\rm gas,G}(t_v)$.  However, a rough estimate of
$t_1$ gives $t_1=0.09 t_v$ (using the spherical collapse model and
assuming that the radius of the system at $t_1$ is equal to the virial
radius).  It therefore seems likely that most of the heating would
occur after $t_1$.

In general, the gravitational contribution to $E_{\rm excess}$ is
difficult to estimate, and would require modelling with hydrodynamic
simulations. It would depend on the total energy injected, and when it
was injected (on average). It would probably be stochastic as well.
This is reminiscent of the `bias parameter' used to relate the
clustering of galaxies to the clustering of DM, suggesting that
perhaps the exact value of $E_{\rm excess}$ can be related to the
energy injected by some bias parameter.

From Fig.~\ref{figensec}, the maximum possible gravitational
contribution would in principle be given by injecting sufficient
energy at $t_a$ to raise $E_{\rm gas}$ to almost zero. Assuming that
the gas so dispersed that $\dpar{\phi}{t}\approx 0$, the gravitational
term in equation~\ref{egasevol} then vanishes. Hence, $E_{\rm
  gas}\approx 0$ at $t_v$, and $E_{\rm excess}$ would be greater than
the energy injected by $E_{\rm max}\approx [E_{\rm
  gas,G}]^{t_a}_{t_v}$.  The axis of $E_{\rm gas}$ in
Fig.~\ref{figensec} has been marked with intervals of $E_{\rm max}$,
according to rough estimates of the absolute values of $E_{\rm gas,G}$
at $t_a$ and $t_v$ (derived in Appendix~C). In this idealized example,
$E_{\rm excess}$ is therefore $\sim 50$ per cent greater than the
energy injected. Such an increase could assist in breaking the
self-similarity of clusters.

\subsection{Work done at the outer boundary} \label{mech2}
If gas in the proto-halo does work on gas outside, we would expect
this to reduce $E_{\rm gas}$, and vice versa. Conceptually this is
quite simple, but we need to follow the gas in more detail than
before.  We introduce the term `proto-gas halo' to refer strictly to
the gas which eventually forms the gas halo. (Thus it does not include
gas that is converted into stars or BDM before virialization. We do
not explicitly account for gas that is recycled from stars, which
should be only a small fraction of the intracluster medium.) The mass
of the proto-gas halo is thus constant with time.

To distinguish this from our earlier discussion, we introduce $e_{\rm
  gas}$ to give the total specific energy at a position comoving with
the gas:
\begin{equation} \label{eegas}
  e_{\rm gas}=\frac{3kT}{2\mu m_{\rm H}} + \half\vel^2 + \phi.
\end{equation}
Integrating this over the mass of the proto-gas halo gives the total
energy, $\int e_{\rm gas}\d m=M_{\rm gas} E_{\rm gas}$. In Appendix B,
we show that
\begin{equation} \label{eegasevol}
  {\d\over\d t}\int e_{\rm gas} \d m = 
    \int (-P\vel+\stress\vel) \cdot\d{\bf A}
    + \int \left( \rho_{\rm g} \dpar{\phi}{t} + \Gamma \right) \, \d V,
\end{equation}
The only change from equation~\ref{egasevol} is the additional surface
integral, in which $P$ is the gas pressure, $\stress$ is the viscous
stress tensor and $\d{\bf A}$ is a vector element of surface area.
The surface integral gives the rate of work done by the proto-gas halo
on other gas. The viscous term is almost certainly negligible for our
purposes, so we assume that it vanishes. The work done
at the outer boundary of the proto-gas halo is then a
straightforward integral of $P \d V$.

Part of the motivation for estimating this is because, if the gas is
heated when it is diffuse, then it is conceivable that the work done
by compressing the proto-gas halo would boost the final excess
energy (due to the increased pressure). Note that it is the change in
work done as a result of heating that we are interested in. Since
hydrodynamic simulations which do not include non-gravitational
processes result in almost self-similar X-ray clusters, we can be
assured that any work done does not prevent them from following a
self-similar energy
equation such as (\ref{energy}).  
For simplicity, we shall consider the total work done after the
turnaround time, to see if this effect can increase $E_{\rm excess}$.

We expect most of the work to occur on those parts of the outer
boundary which form the ends of filaments and, possibly, the edges of
sheets, because density and temperature are highest at these surfaces.
Although the rest of the outer boundary has a much larger area, we
shall assume that the pressure there is so small that the
work done there is no more than that at the end of filaments.
For filaments, the volume swept out by the end surfaces should be
comparable to the volume of the filaments. This is because infall
occurs along the filaments in general. Using the spherical
collapse model for comparison, the volume of the sphere at turnaround
is $8 V_{200}$, where $V_{200}$ is the volume of the virialized halo.
Let there be an effective pressure $P_{\rm eff}$, then the work done
on the proto-gas halo between the turnaround time and virialization
is $7 f_{\rm eff} V_{200} P_{\rm eff}$, where $7 f_{\rm eff} V_{200}$
is the effective volume swept out by the said surfaces.
Letting $P_{\rm eff}=\rho_{\rm g,eff} kT_{\rm
  eff}/(\mu m_{\rm H})$, where we have defined an effective density and
temperature, the work done is then given by
\begin{equation}
  7 f_{\rm eff} V_{200} {\rho_{\rm g,eff} kT_{\rm eff} \over \mu
  m_{\rm H}} =
     7 f_{\rm eff} {\rho_{\rm g,eff} \over \overline{\rho_{\rm g}}}
     kT_{\rm eff} \left(M_{\rm gas}\over \mu m_{\rm H}\right),
\end{equation}
where $\overline{\rho_{\rm g}}$ is the mean density of the virialized
gas halo and $M_{\rm gas}/\mu m_{\rm H}$ is the number of particles in
the gas halo. It follows that the contribution to $E_{\rm gas}$ is
given by
\begin{equation}
  7 f_{\rm eff} {\rho_{\rm g,eff} \over \overline{\rho_{\rm g}}}
  \left(kT_{\rm eff}\over \mbox{keV}\right) \mbox{keV/particle}.
\end{equation}

The volume filling factor of filaments, which we use to approximate
$f_{\rm eff}$, naturally depends on the threshold density above which
we define our filaments.  From hydrodynamic simulations of the IGM in
a Cold Dark Matter $\Omega=1$ cosmology (Zhang et al.
1999)\nocite{zman99}, threshold overdensities of about 1 to 5
(relative to the background baryon density) result in filamentary
structures, but higher than $\sim 10$, the structures obtained become
dominated by knots rather than filaments. Most of the filamentary
structures also appear to be in place by $z=5$, and exhibit mild
evolution after that \cite{zman99}.  We shall use a fiducial value of
$f_{\rm eff}\sim 0.01$ and a fiducial overdensity of 10. In an
$\Omega=1$ universe $\overline{\rho_{\rm g}}$ is 200 times the
background density.  In the simulation, the filaments have a typical
temperature of $\sim 10^{-3}$ keV.  We thus obtain a fiducial value of
$7\times 0.01\times 0.05\times 10^{-3}=3.5\times 10^{-6}$ keV/particle
for the work done in the absence of heating. This is clearly
negligible.

If all of the gas in the filaments is heated strongly to a temperature
of $\sim 1$ keV, then the gas halos within would also be flushed out.
This would momentarily increase the effective density of the
filaments.  Substituting the values ${\rho_{\rm
    g,eff}/\overline{\rho_{\rm g}}}=1/3$ and $kT_{\rm eff}=1$ keV, the
work done becomes 0.02 keV/particle. In reality, the gas would
continue expanding out of the filaments, so that the volume filling
factor $f_{\rm eff}$ would increase. Assuming that the gas expands
adiabatically, $\rho_{\rm g,eff} T_{\rm eff}\propto P_{\rm eff}\propto
1/f_{\rm eff}^\gamma$, where $\gamma=5/3$. This suggests that the
actual work done would be less than the above estimate, and therefore
$\ll 1$ keV/particle. The caveat is that we have not accounted for gas
that is more diffuse than the filaments, which may also be heated to
$\sim 1$ keV/particle.

\subsection{Work done on hot gas that cools}
In this and the following section, we consider how the conversion of
gas into stars and BDM {\em inside} the proto-halo may affect the
total energy of the gas remaining.

If a progenitor halo contains hot gas which cools out, then work may
be done by the proto-gas halo on the gas that cools. (In cold
collapses this should be very small, as the gas is in general not
pressure-supported.)  This would reduce the total energy of the
proto-gas halo. However, the gas remaining started out at larger
radii; therefore it had a higher than average potential energy before
cooling started. This effect is discussed more generally in the next
section, and it works in the opposite direction.  The net effect can
be investigated with a spherical hydrodynamic simulation of a hot gas
halo which cools.

Here, we describe a simple way to obtain an upper limit on the work
done, using our simulations.
Suppose the proto-gas halo has an inner surface or `bubble' that lies
inside some progenitor halo that contains hot gas (the
`bubble' is likely to quite irregular). Let $\rhog$ and $T$ be the
density and temperature at this surface.  Then the work done as the
bubble shrinks is $\int P \d V=\int kT/(\mu m_{\rm H})\, \rhog \d V$.
We shall assume that the gas halo is isothermal. Now, $\rhog \d V$ at
the bubble wall is smaller than the mass of the corresponding gas that
cools out, because the latter had an initial density greater than
$\rhog$. It follows that if we replace $\rhog \d V$ in the integral
with $\d m$, the mass of gas that cools, then we would
overestimate the work done. Hence $kT m_{\rm BDM}/(\mu m_{\rm H})$ gives
an upper limit on the total work done, where $m_{\rm BDM}$ is the mass
of hot gas converted into BDM in our simulations. (Note that it does
not actually matter whether the gas is converted into BDM, stars, or a
cold disk.)

However, if {\em all} of the hot gas in the progenitor halo cools out,
then the `bubble wall' must lie outside the gas halo, where the
pressure is probably negligible. This suggests that we should not
count such cases at all.

Over the history of the proto-gas halo, the total work done on hot gas
that cools is thus $<\sum kT m_{\rm BDM}/(\mu m_{\rm H})$, where the
summation is made over all progenitor halos that did not cool out all
of their hot gas. The reduction in $E_{\rm gas}$ is therefore less
than
\begin{equation}
  \left(1\over M_{\rm gas}\right) \sum m_{\rm BDM}\left(kT\over
  \mbox{keV}\right) \mbox{keV/particle}.
\end{equation}
We computed this quantity using Model B (\ie only isothermal
gas profiles) and both sets of parameters given in Table~\ref{tabte}.
For small clusters ($T\approx 2$ keV), we obtain around
0.25 keV/particle, with a scatter of 50 per cent each
way. For large clusters ($T\approx 10$ keV) the upper limit is
about double this. The two simulations gave similar results.

[We note that the bubble is just an imaginary surface for
separating different subsets of gas. If heating (in the form of
$\Gamma$) occurs inside a bubble, gas outside can still be heated via
the surface term in equation~\ref{eegasevol}. For most purposes, the
distinction is best ignored.]

\subsection{The effect of gas removal} \label{mech4}
Having developed the machinery to follow clumps of gas individually,
it is natural to ask whether the spatial distribution of
the proto-gas halo can itself result in excess energy. This becomes
clear if we consider the proto-gas halo at very early times.  Its
outer boundary is then almost spherical, but it would contain many
`bubbles' inside, as described above. If the bubbles occur
preferentially towards the centre of the sphere, then the gas would
have positive excess energy, because fractionally more gas would
be found at larger radii and higher potentials than in
a uniform distribution. Again, we are comparing to the case without
non-gravitational processes, for which the proto-gas
halo is just a uniform sphere at very early times.
In section~\ref{selsec}, we suggest how a positive excess energy can
occur in this way, and make simple estimates of its magnitude.

To estimate the excess energy, it is easier to make comparisons when
the halo has virialized, because the time evolution of $\int e_{\rm
  gas} dm$ is complicated. Consider a virialized gas halo obtained
without cooling: only a subset of its gas particles would remain in
the gas halo if the system were evolved with cooling included. If this
subset has a more extended distribution than the entire
gas halo, then the subset would have a positive excess specific
energy. Assuming that the gas is isothermal for simplicity, $E_{\rm
  excess}$ can be estimated by comparing $E_{\rm gas}$ for the subset
with that for the entire gas halo.

We note that the above method actually overestimates $E_{\rm excess}$,
for it does not account for the work done on the cooling gas, and it
probably overestimates the gravitational contribution in the following
way. Since we compute $E_{\rm gas}$ after virialization, the evolution
of $\int e_{\rm gas} \d m$ for the subset of gas particles is already
accounted for.  Since the subset is more extended than the whole,
there is likely to be a positive gravitational contribution.  However,
in reality gas belonging to the subset would gradually fall to smaller
radii to replace cooled gas. The method does not account for this and
therefore overestimates the gravitational contribution. If a
hydrodynamic simulation of a cluster is performed with cooling, all
these effects would be naturally accounted for. In this case, $E_{\rm
  excess}$ could be computed exactly by comparing with the same cluster
evolved without cooling.

\section{Breaking the self-similarity of clusters} \label{breaksec}

In section~\ref{clustersec}, 
we showed that excess energies of about 1 keV/particle
or more are required to match the properties of X-ray clusters.
However, we found that our model generates only $\sim 0.1$
keV/particle from supernova heating. Nevertheless, the excess energy
deduced from the iron abundance of X-ray clusters can be as high as 1
keV/particle (WFN98).  To obtain this result, we
made two crucial assumptions: that most of the iron originated from
type II supernovae (SNII), and that a large fraction of the supernova
energy---we assumed $4\times 10^{50}$ erg per supernova---is
retained.

Unfortunately, the first assumption is already in doubt. A recent
analysis suggests that SNIa supply 30--90 per cent of the iron in
clusters, depending on the supernova model \cite{ns98}. Recall that
the same amount of iron contributed by SNIa corresponds to $\sim 10$
times less energy.  As for the supernova energy that is retained,
\citeN{tgjs98} have made a systematic study of supernovae exploding in
cold gas (1000 K) in a range of gas densities and metallicities. They
find that in the late stages of evolution, the supernova remnants have
total energies of about $(9$--$30)\times 10^{49}$ erg (they assumed
initial energies of $10^{51}$ erg per supernova).  We note that if the
supernova rate is sufficiently high that remnants overlap before going
radiative, then the heating efficiency may be higher in reality. We
are therefore unable to rule out supernovae as the source of the
required energy, based on the present data. However, it is our opinion
that this scenario is only marginally acceptable.

The purpose of this section is to move beyond the confines of our
model, and discuss other possible approaches to breaking the
self-similarity of clusters.

\subsection{Supernova heating} \label{snheatsec}

Assuming that all the excess energy can be provided by supernovae, we
consider the basic properties that such a model would need to have.
First of all, it is clear that a large fraction of the iron
in clusters would have to come from SNII. To obtain enough supernovae,
most of the stars observed in present day clusters would need to be
formed with a flattened IMF: for example, boosting the standard
supernova rate by a factor of 5, and assuming a gas-to-stellar mass
ratio of 15 (the same parameters as in section~\ref{galsec}), gives an
iron abundance of $Z_{\rm SNII}=0.15$\Zs\ provided all of the iron is
deposited in the intracluster gas. This corresponds to 1 keV/particle
if we set \esn=1.8. (Since this is already very high for \esn, we
would not want $Z_{\rm SNII}$ to be much lower.) Note that practically
all SNII would have to have such a high heating efficiency, therefore
most star-forming galaxies would have to be involved in the heating
process. 

We showed in section~\ref{galsec} that the main obstacle to obtaining
higher excess energies in our model was the assumption that gas is
ejected at the escape velocity of host halos. In order for gas heated
by supernovae to escape from a halo with much greater than the escape
energy, it needs to find a clear path out of the halo.  Unfortunately,
this is difficult if the site of star formation is surrounded by a hot
gas halo, and a continuous infall of cooling gas may also be
problematic.  If a clear path is {\em not} found, then gas surrounding
the site of star formation would be heated gradually; it would leave
the heat source as soon as it had sufficient energy to escape the
halo, hence it would be ejected with no more than the escape energy.
Since halos $\ga 10^{12}$\Ms\ generally contain hot gas, we shall
suppose that the heating occurred in less massive halos. In addition,
we show in a separate paper \cite{wfn99x}
that most of the gas associated with halos in the
range $\sim 5\times 10^{12}$--$10^{14}$\Ms\ must lie outside their
virial radii, therefore it is clear that at least some of the
heating must have occurred in less massive halos. The gas would
therefore be ejected with excess energies $\ga 10$ times the binding
energy of these halos (see Fig.~\ref{figbe}). To do this, supernovae
would need to carve out `chimneys' in the surrounding gas, for the hot
gas to escape from. This could be made easier by delaying star
formation until most of the gas has settled into a cold disk, \eg by
magnetic pressure, turbulence and/or angular momentum support.

To keep radiative loss to a minimum, gas needs to be rapidly heated to
very high temperatures ($\sim 1$ keV) and then ejected. Gas that is
not ejected must not receive much of the energy, as the cooling times
of galaxies are relatively short.  There is also the problem of
dilution: if only a certain fraction of the intracluster medium is
heated in this way, then gas would need to be ejected with
correspondingly higher excess energies.

A side effect of this scenario is that the ability of supernovae to
regulate star formation would be greatly diminished. Since most of the
energy must be channelled into gas that is ejected, the main role of
supernovae is to regulate the quantity of cold gas that is left in the
halo.  However, for the same amount of supernova energy, the amount of
gas that is ejected is $\la 1/10$ of that in a more conventional model
which assumes gas is ejected at about the escape velocity. Therefore,
another more effective regulator of star formation would be
required. Otherwise, the bulk of star formation would occur in the
smallest halos, and more massive halos would become very
gas-deficient. One possibility would be to assign a long
time-scale for star formation in cold gas, following other SAMs
\cite{kwg93,cafnz94,sp98}, though the published time-scales would need
to be increased.

Is it possible for supernova heating to continue in the hot gas halos
of groups? Observations suggest that 10--20 per cent of cold gas
deposited in cooling flows may form stars \cite{cga98}. In addition,
the binding energies of halos $\ga 10^{13}$\Ms\ in mass are around 1
keV/particle or more, so it would be possible to reach the required
excess energies in these halos without ejecting their gas.  We shall
briefly discuss some of the difficulties with this model. Firstly,
groups are gas-poor compared to clusters \cite{wfn99x}, so it is
unclear how much gas would be available to form stars, especially as
the cooling times are increased by the low gas density. Secondly,
gas cools and forms stars gradually in a cooling flow, so that
supernovae would heat surrounding gas which is in the process of
cooling. It is therefore unclear how efficiently supernovae can heat the gas
that does not cool, since the heating may simply slow down the cooling
flow.

In any case, the model described is already tightly
constrained by the present data, so it can be tested in several ways
with future observations.  Spatially-resolved spectral analysis will
allow us to properly measure the average metallicity of intracluster
gas.  Most of the present measurements are emission-weighted, which
would overestimate the average metallicity if a negative metallicity
gradient is present. Better estimates of the SNIa contribution to the
iron abundance may also rule out the above model. Our own results
suggest that if X-ray clusters with $T\sim 2$ keV turn out to be
isothermal, then their excess energy should be about 2 keV/particle,
instead of 1 keV/particle (see Fig.~\ref{figetagamtab}).
Spatially-resolved temperature and density profiles would therefore
further constrain the energy requirements and therefore the models
that are allowed.

The above discussion may be altered if hypernovae releasing $\sim
10^{52}$ erg each \cite{iwamo98} were common. Since the progenitors of
hypernovae are believed to be stars of mass $\ga 40$\Ms, such a
scenario would still require an IMF strongly biased towards very
massive stars.

\subsection{Pre-collapse gas at high entropy} \label{ponmansec}

Thus far, we have used the total energy of a gas halo as the main
constraint on its structure.  In this section we shall discuss a different
constraint, namely the gas entropy, which we measure with the
quantity $s=T/n_e^{2/3}$.

It was proposed by \citeN{kaise91} and \citeN{eh91} that a better match
to the $L_{\rm X}-T$ relation could be obtained if
the IGM was `preheated' to a high entropy prior to collapse of the
gas.  \citeN{nfw95} used this method in hydrodynamic simulations of
three clusters, using a gas fraction of 0.1 in an Einstein-de-Sitter
universe. By giving all gas particles a uniform high entropy at a
redshift of $z=3$ (no radiative cooling was included), they were able
to obtain clusters that closely followed $L_{\rm X}\propto T^3$.

More recently, Ponman et al.\ (1999) have measured the entropy of gas in
clusters at one tenth of the virial radius (to avoid possible cooling
flow regions) and found that the entropies measured in poor clusters
and groups were higher than predicted assuming self-similarity.
Instead, the entropies appeared to settle on a lower limit or `floor'
given by $T/n_e^{2/3}\sim 100 h^{-1/3}$ keV~cm$^{2}$. This suggested
that perhaps all of the gas had been preheated to this entropy, so that
outside any cooling region, the entropy would have at least this value
(since shock heating always increases the entropy). \citeN{bbp99}
investigated this idea by assuming that the preheated gas evolves
adiabatically.  Using an initial entropy consistent with the observed
`entropy floor', they found that the isentropic model could fit the
properties of groups ($T\la 1$ keV) but could not match the properties
of clusters. This was attributed to the need for accretion shocks to
raise the entropy further in clusters.

We have stressed that clusters need to have sufficient excess energy
in order to match the $L_{\rm X}-T$ relation. We therefore argue that
preheating the gas to an entropy floor alone would not solve the
problem unless the excess energy is sufficiently high. However, this
may turn out to be a superfluous point, since creating an entropy
floor probably requires large amounts of energy anyway (see below). We
also note from Fig.~\ref{figetagamtab2} that the gas profile which
requires the least excess energy to match the data also has the highest
entropy near the centre. Therefore it clearly helps if we try to
raise the entropy as high as possible.

The energy required to raise the entropy to a certain level depends
very much on the density of the gas. Since the IGM is very diffuse in
places, it seems that a relatively high entropy can be achieved with
very little energy. However, the difficulty is heating most of the gas
in the proto-cluster in this way, especially the gas which eventually
forms the core of the cluster. This is because the minimum density
experienced by the gas is limited by the overdensity that led to the
cluster in the first place (as well as smaller scale density
fluctuations).  The spherical collapse model gives a simple
illustration of this constraint.

To estimate the `advantage' of heating the gas at low density,
we need to compare the density at the time of heating to the final
density of the gas, \ie we need to estimate the compression ratio.
Using the spherical collapse model, the turnaround radius of the
sphere is twice the final virial radius, so that the mean density of
the sphere has a minimum equal to 1/8 of the mean density of the
virialized halo. This simple model suggests that adiabatic compression
can increase the temperature of preheated gas by a factor of 4 at
most.

Alternatively, we can compare the minimum
density obtained above to a fiducial density of $n_{\rm e} = 10^{-3}
\rm\ cm^{-3}$ near the centre of a cluster (above which cooling can
significantly modify the entropy during the life of the cluster).  The
mean gas density at turnaround for a halo that collapses at time $t$
is given by $200 f_{\rm gas} / (48 \pi G t^2)$, where $f_{\rm gas}$ is
the gas fraction. This implies an electron density of $n_{\rm
  e,min}=2.0\times 10^{-5} f_{0.2}/t_{10}^2$ cm$^{-3}$, where
$t_{10}=t/(10^{10}$ years) and $f_{0.2}=f_{\rm gas}/0.2$.  Thus the
temperature increase when this gas is compressed to density $n_e =
10^{-3} \rm\ cm^{-3}$ is a factor of $13 t_{10}^{4/3}
(n_{-3}/f_{0.2})^{2/3}$, where $n_{-3}=n_e/(10^{-3}$ cm$^{-3})$.  In
reality, this factor is much reduced by clumping of the gas into
filaments and sheets by the turnaround time, so that typical value of
$n_{\rm e,min}$ should be much higher than we have estimated. In
addition, the central region of the cluster, with roughly the fiducial
density of $n_{\rm e}= 10^{-3} \rm\ cm^{-3}$, almost certainly virialized
at an earlier time as a less massive halo. If {\em this} virialization
time is used in $t_{10}$ and filamentary nature of the gas at
turnaround is accounted for, then the temperature increase due to
adiabatic compression is only a factor of a few. 

To obtain more precise estimates, we need to use a hydrodynamic
simulation. Assuming that the temperature increase is a factor of 4,
the isentropic profile in Fig.~\ref{figetagamtab2} which requires the
least excess energy has a temperature of just over 3 keV at $0.1
r_{200}$. This would therefore imply a temperature of $(3/4)$ keV before
compression, or a thermal energy of $(9/8)$ keV/particle.

\subsection{Heating by active galactic nuclei} \label{agnsec}

Energetically speaking, the total energy released in the formation of
massive black holes at the centres of galaxies is sufficient to heat
all the baryons in the universe to very high excess energies. However,
the mechanism for injecting this energy into the gas is uncertain:
this may occur through jets and winds, but the energy released in this
form is not well known. On the other hand, the energy released as
radiation is relatively well measured.  

\citeN{ewnb98} have estimated 
the total energy released by black hole formation in the Coma cluster.
They assumed a mass-to-light conversion rate of $\epsilon\approx 0.1$
and roughly the same rate of energy release in relativistic particles
and magnetic fields (as in the jets of radio galaxies).  They
concluded that the total energy released in the latter form was
comparable to the thermal energy of the gas in the Coma
cluster. Therefore, if all of this energy was injected into the gas,
it could significantly modify the gas distribution of the cluster.

It is possible to make a similar estimate by averaging over all the
baryons in the universe.  From the observed luminosity density of AGN,
the total mass density of black holes in the universe can be
determined.  Using the X-ray background intensity at 30 keV,
\citeN{fi98} obtain a range of 6--$9\times 10^5$\Ms\ Mpc$^{-3}$ for
the black hole density. [This is higher than earlier estimates
\cite{solta82,ct92}, which used optical counts of AGN. It is
likely that these counts suffered from strong intrinsic absorption
\cite{fbai98}.] Assuming a mass-to-light conversion rate of 0.1 and a
black hole density of $6\times 10^5$\Ms\ Mpc$^{-3}$, the total energy
radiated by AGN is then $6.4\times 10^{58}$ erg Mpc$^{-3}$.  If the
same amount of energy is available in relativistic particles and
magnetic fields, and it is divided uniformly over all the baryons in
the universe, we would obtain an energy injection of 3.7
keV/particle. As before, we have assumed $\Omega_b=0.08$ and $h=0.5$.
This amount of heating would therefore be more than enough to break the
self-similarity of clusters.

On the downside, we note that only about 10 per cent of AGN have
observed radio jets. If such jets provide the only mechanism for AGN
to heat surrounding gas, then the estimated excess energy would be
correspondingly reduced. However, it is possible that radio-quiet
quasars may also heat surrounding gas through outflows of thermal gas
or poorly-collimated `jets' of radio-emitting plasma
\cite{kunci99,fabia99}.

The advantage of this form of heating over supernova heating is that
it need not be intimately connected with the process of star
formation. By obtaining the required energy from AGN, supernovae would
be able to perform their usual role as regulators of star formation
(see above). In addition, since an AGN is a single powerful source of
energy, the gas being heated is more likely to be raised quickly to
a very high temperature ($\ga 1$ keV). In this case, the cooling times
would be comparable to those of X-ray clusters and radiative loss
would be minimised.

\subsection{Preferential removal of gas} \label{selsec}

As explained in section~\ref{mech4}, it is possible that the removal
of cooled gas can result in an excess specific energy in the
gas that remains to form the intracluster medium.

The excess energy can be estimated from the subset of gas particles,
in a cluster evolved {\em without} non-gravitational processes, which
would remain in the gas halo if radiative cooling is included. If
the subset has a more extended distribution than the entire gas halo,
then a positive excess energy would result. This would occur if gas at
smaller radii had a higher probability of cooling out than gas at
larger radii.

Such a scenario may occur as follows. Theory predicts
that the first halos of a given mass to collapse should be much more
strongly clustered than the background density distribution (Kaiser
1984).  For instance, the large-scale over-densities that created
present-day clusters also raised the overall density of smaller-scale
fluctuations, so that the first galaxy halos to collapse had a high
probability of being associated with future clusters. The above has
been used to explain the strong clustering of `Lyman-break galaxies'
(LBG) observed at $z\sim 3$ \cite{steid98,adelb98,giava98}, where good
agreement with theoretical predictions have been obtained if the
typical LBG is associated with a halo of mass $\sim 10^{12}$\Ms.
N-body simulations show that the densest peaks in the distribution of
LBGs are likely to be the progenitors of future clusters
\cite{gover98,wgpbd98}.  If we make the reasonable assumption that the
large-scale over-density that led to a cluster was highest near the
centre, then it seems likely that the LBGs would form preferentially near
the centre of the cluster.  Naturally, as more of the proto-cluster
goes non-linear, galaxies would become more uniformly distributed in
the proto-cluster.  Nevertheless, the first sub-halos of a given mass to
collapse also have the highest mean gas density, so that they have the
shortest cooling times. Hence gas is more likely to cool out, and be
removed, near the centre of the cluster.

To obtain an upper bound on the excess energy obtainable in this way,
we modelled the virialized cluster (in the no-cooling case)
with singular isothermal spheres ($\rho\propto r^{-2}$) for both the
gas and dark matter.  Assuming a primordial baryon fraction of 0.27, a
cluster gas fraction of 0.17 is obtained if we remove all of the gas
inside a radius of $(10/27)r_{200}$ in the above gas distribution
(recall that this amount of cooling was obtained in our simulations).
The difference in $E_{\rm gas}$ before and after the gas is removed
thus gives the excess energy. Since the gas is isothermal, it is only
necessary to calculate the gravitational term in $E_{\rm gas}$, for the
thermal terms cancel when we take the difference.  The result is
$E_{\rm excess}=(10/17)\ln(10/27) GM_{\rm tot}/r_{200}=0.58 GM_{\rm
  tot}/r_{200}$, where $M_{\rm tot}$ is the total mass of the halo.
For the cluster displayed in Fig.~\ref{figetagamtab} (which has a
virial radius of 1.46 Mpc), this gives an excess energy of 1.4
keV/particle.

In reality the gas removed must be more extended than assumed above.
Removing a uniform fraction of gas at each radius naturally leads to
no excess energy. If we model the more general case by removing the
gas in two component: a `uniform' component, followed by all the gas
inside a radius of $fr_{200}$, then we get
\begin{equation}
  E_{\rm excess}=-{f\ln f\over 1-f} {GM_{\rm tot}\over r_{200}}.
\end{equation}
For example, if half of the gas removed 
in the uniform component, then $f=5/(27-5)=5/22$. This gives $E_{\rm
excess}=0.44 GM_{\rm tot}/r_{200}$, or 1.0 keV/particle for the above
cluster. Increasing the uniform component to 3/4 of the gas removed,
so that $f=2.5/19.5$, gives $E_{\rm excess}=0.30 GM_{\rm tot}/r_{200}$
or 0.7 keV/particle.

In a separate paper \cite{wfn99x},
we show that groups are even more strongly affected by
heating than clusters, so that most of their gas is outside their
virial radii. It does not seem possible for cooling alone to explain
this phenomenon. Therefore, the above mechanism
would have to be supplemented by heating in the conventional sense.

\subsection{The bottom line}

Of the three main methods discussed, supernova heating appears only
marginally acceptable based on current data, and requires a much
higher heating efficiency than is commonly assumed. Preferential
cooling also struggles to provide sufficient excess energy, and would
not be able to explain our results for groups. Since it is possible
for AGN to provide more than enough energy, this would be our
preferred choice. However, the actual heating mechanism in this case is
uncertain. It remains possible that all three mechanisms
contribute to the excess energy.

\section{Conclusions} \label{summary}

We have constucted a self-consistent semi-analytic model which follows
the excess energies resulting from supernova heating and radiative
cooling, and modifies newly-collapsed gas halos accordingly. The gas
profiles of virialized halos are selected from a 2-parameter family of
polytropic gas profiles in NFW (1997) potential wells.

In the absence of non-gravitational heating or cooling, the gas halos
of model clusters are approximately self-similar, in agreement with
the results of hydrodynamic simulations. In particular, their bulk
properties follow self-similar scaling laws such as $L_{\rm X}\propto
T^2$. The model was then normalized by matching to the largest
observed X-ray clusters, as these are least affected by
non-gravitational heating.

Four contrasting `heating models' were used to investigate the excess
energy required to match X-ray cluster data. Each heating model
represented a different way of modifying gas profiles in the presence
of heating.  In addition we investigated the excess energy available
from supernova heating in our model, and discussed effects our model
could not account for which may possibly contribute to the excess
energy of gas halos. In the last section, we discussed other
approaches to obtaining the required excess energy, including a
significantly modified model for supernova heating, heating by AGN,
and the removal of gas at low potentials.

We summarize our main conclusions below:

\begin{itemize}
\item The semi-analytic model is able to reproduce the observed
  $L_{\rm X}-T$ relation, temperature function, luminosity function
  and mass deposition rate function, provided the simulated X-ray
  clusters are given excess energies of $\sim 1$ keV/particle in order
  to break their self-similarity.
\item The excess energies required by each of the four heating models
  to match the observed $L_{\rm X}-T$ relation lie in the range
  1.8--3.0 keV/particle. By analysing a fiducial cluster with $T\approx
  2$ keV, we find that the minimum excess energy required is about 1
  keV/particle when all the available gas profiles are considered (the
  winning profile in this case is isentropic).
  We note that other authors require similar amounts of heating
  \cite{pen99,loewe99}.
\item If the process that produces the excess energy ejects gas in
galactic winds at the escape velocity of the host halo (as assumed by
our model), then the resulting excess energies in halos of all masses
follow a distinct pattern. This is largely determined by the binding
energies of halos and the halo merger tree. The excess energies are
therefore not sensitive to parameters such as the efficiency of
supernova heating, \esn.
\item In this case, the resulting excess energies in clusters are only
  $\sim 0.1$ keV/particle, an order of magnitude less than the required
  amount.
\item If the gas distribution is made more extended by a high level of
energy injection before the cluster virialized, then a positive
`gravitational contribution' to the excess energy is likely. This may
help to ease the energy requirements and will need to be investigated
with hydrodynamic simulations.
\item Of the approaches discussed in section~\ref{breaksec} for
obtaining the required excess energy: more than enough energy is
available from AGN, supernova heating is only marginally acceptable,
and preferential cooling struggles to provide sufficient excess
energy. However, it remains possible that all three mechanisms
contribute to the excess energy of X-ray clusters.
\end{itemize}

It seems likely that similar excess specific energies to those in
clusters also occur in groups \cite{wfn99x}, in which case a large
fraction of the gas that belongs to groups would be outside their
virial radii.  This may explain their steeper $L_{\rm X}-T$ relation
(see also Balogh et al. 1999).  

Future measurements of the gas density and temperature profiles of
groups and small clusters should clarify these issues, and place much
stronger constraints on the excess energy in low-temperature clusters.

\section*{Acknowledgements}
KKSW thanks Vince Eke, Stefano Ettori, Martin Haehnelt, Fraser Pearce,
Clovis Peres, Martin Rees, Joop Schaye and Tom Theuns for helpful
discussions. KKSW is grateful to the Croucher Foundation for financial
support. ACF thanks the Royal Society for support.

\bibliography{mnrasmnemonic,paper,bookastrophys}
\bibliographystyle{mnras}

\appendix
\section{Formulae for modelling the gas processes}
The equations used to model the gas processes described in
section~\ref{briefsec} are given below, along with formulae for some
observed quantities. Where required, we assume polytropic gas profiles
in NFW potential wells, as derived in sections~\ref{nfwsec} and
\ref{familysec}.  We remind the reader that the gas profiles used in
the formulae are notional, as explained in section~\ref{coldhot}. To
denote radius we use $r$ and $x$ interchangeably, where $r$ is
the physical radius and $x=r/r_s$.

We always calculate the quantities $L_{\rm X}$, $\dot{M}$,
emission-weighted temperature, and cooling flow power at a redshift of
zero, thus any evolution in these quantities over the life of a halo
is accounted for.

\subsection{The extent of cold gas, $x_{\rm cf}$} 

When a new halo forms, the ratio between the cooling time and the
free-fall time to the centre of the halo, $\tau=t_{\rm cool}/t_{\rm
  ff}$, determines whether gas is able to form a hot hydrostatic
atmosphere.  A hot gas halo forms when $\tau>\tau_0$, where $\tau_0$
is a parameter of the model.  If $\tau<\tau_0$ then the gas remains
cold in general (as virialization shocks would be radiative and any
heating would be transitory). When $\tau$ is greater or less than
$\tau_0$ everywhere, the above criteria are simple to apply.
Otherwise, if $\tau$ increases monotonically with radius, then there
is a unique radius, $x_{\rm cf}$, where $\tau=\tau_0$. Gas inside of
$x_{\rm cf}$ is then classified as cold and the remaining gas forms a
hot gas halo. In all cases, our model requires there to be one radius
$x_{\rm cf}$ which lies in the range 0 to $c$, such that gas inside
$x_{\rm cf}$ is classified as cold and that outside, hot. Since
$\tau(x)$ is comparable to $\tau_0$ in a narrow range of halo
masses (corresponding to normal galaxies), the variation of $\tau$ with
radius is of concern only for this mass scale. For this reason we will
only discuss in detail gas profiles used in section~\ref{galsec} (\ie
those belonging to Models A and B). We first derive the general
expression for $\tau(x)$ before considering less well-behaved cases.

The cooling time of gas is given by
\begin{equation}
  t_{\rm cool}={3\over 2}{\rho_{\rm g}
                   kT/\mu m_{\rm H}\over n_e n_{\rm H} \Lambda(T)},
\end{equation}
where $\rho_{\rm g}$, $T$, $n_e$ and $n_{\rm H}$ (the electron and
hydrogen number densities respectively) are all functions of $r$. The
three densities are simply proportional to each other.  The cooling
rate is given by $n_e n_{\rm H} \Lambda(T)$, where $\Lambda(T)$ is the
cooling function. We use the cooling function of \citeN{bh89}, which
depends on metallicity as well as temperature. We assume that the
metallicity of every gas halo is constant with radius. A simple
estimate of $t_{\rm ff}$ is obtained by computing the free-fall time
for a test particle to reach the centre of a sphere of uniform
density:
\begin{equation}
  t_{\rm ff}=\sqrt{3\pi\over 16G\rho_{\rm tot}},
\end{equation}
where $\rho_{\rm tot}$, the total density of the halo at the radius
concerned, has been substituted for this density. (The formula given
is a factor of $\sqrt{2}$ greater than that for a collapsing sphere of
uniform density.) This method does not account for the increased
$\rho_{\rm tot}$ towards the centre of the halo, hence it is a slight
overestimate.

It follows that
\begin{equation}
  \tau={t_{\rm cool}\over t_{\rm ff}}=
 \left({3\over 2}\sqrt{16G\over 3\pi}{\rho_{\rm g}^2\over n_e n_{\rm H}}\right)
    {\alpha\over\eta_{200}} {T\over T_{200}}
    {\rho_{\rm tot}^{1/2}\over \Lambda(T)\rho_{\rm g}},
\end{equation}
where we have used the expressions $\eta_{200}=\alpha\mu m_{\rm
  H}/(kT_{200})$ and $\alpha=4\pi G\rho_s r_s^2$, the latter being the
characteristic potential of the NFW profile. We assume a primordial
composition of 0.9 hydrogen to 0.1 helium by {\em number}, which gives
$\mu=0.619$ and $\rho_{\rm g}^2/(n_e n_{\rm H})= 1.707 m_{\rm H}^2$.
Expanding $\rho_{\rm tot}(x)$ and $\rho_{\rm g}(x)$, we obtain
\begin{equation}
  \tau(x)= 
 \left({3\over 2}\sqrt{16G\over 3\pi}{\rho_{\rm g}^2\over n_e n_{\rm H}}\right)
   {\alpha\rho_s^{1/2}\over\eta_{200}\rho_{\rm g,200}}{1\over \Lambda(T) g(x)},
\end{equation}
where we have defined
\begin{eqnarray}
 \lefteqn{ g(x)=x^{1/2}(1+x) } \nonumber \\
 & &\!\!\times\left[1+{\gamma-1\over\gamma}\eta_{200}
  \left({\ln(1+x)\over x}-{\ln(1+c)\over c}\right)\right]^{{1\over\gamma-1}-1}.
\end{eqnarray}
If $0<x_{\rm cf}<c$, then the equation $\tau(x_{\rm cf})=\tau_0$ is
solved numerically. 

To obtain $x_{\rm cf}$, the model follows an algorithm which first
determines whether or not $\tau(x)$ is `well-behaved'. This is done by
first approximating it as proportional to $1/g(x)$. Although $\Lambda(T)$ is
a complicated function when considered over several decades of
temperature, the amount that $T$ can vary in a given halo is limited.
The steepest temperature profile we use that may be of concern is
given by $\gamma=5/3$ and $\eta_{200}\approx 10$, which is used in
Model A. Here the temperature rises by about a factor of 3.5 from
$r_{200}$ to the centre.  In general, the temperature range in a halo
is much smaller, so that the mean variation of $\Lambda(T)$ in halos
is not large.

\begin{figure}
\centerline{\psfig{figure=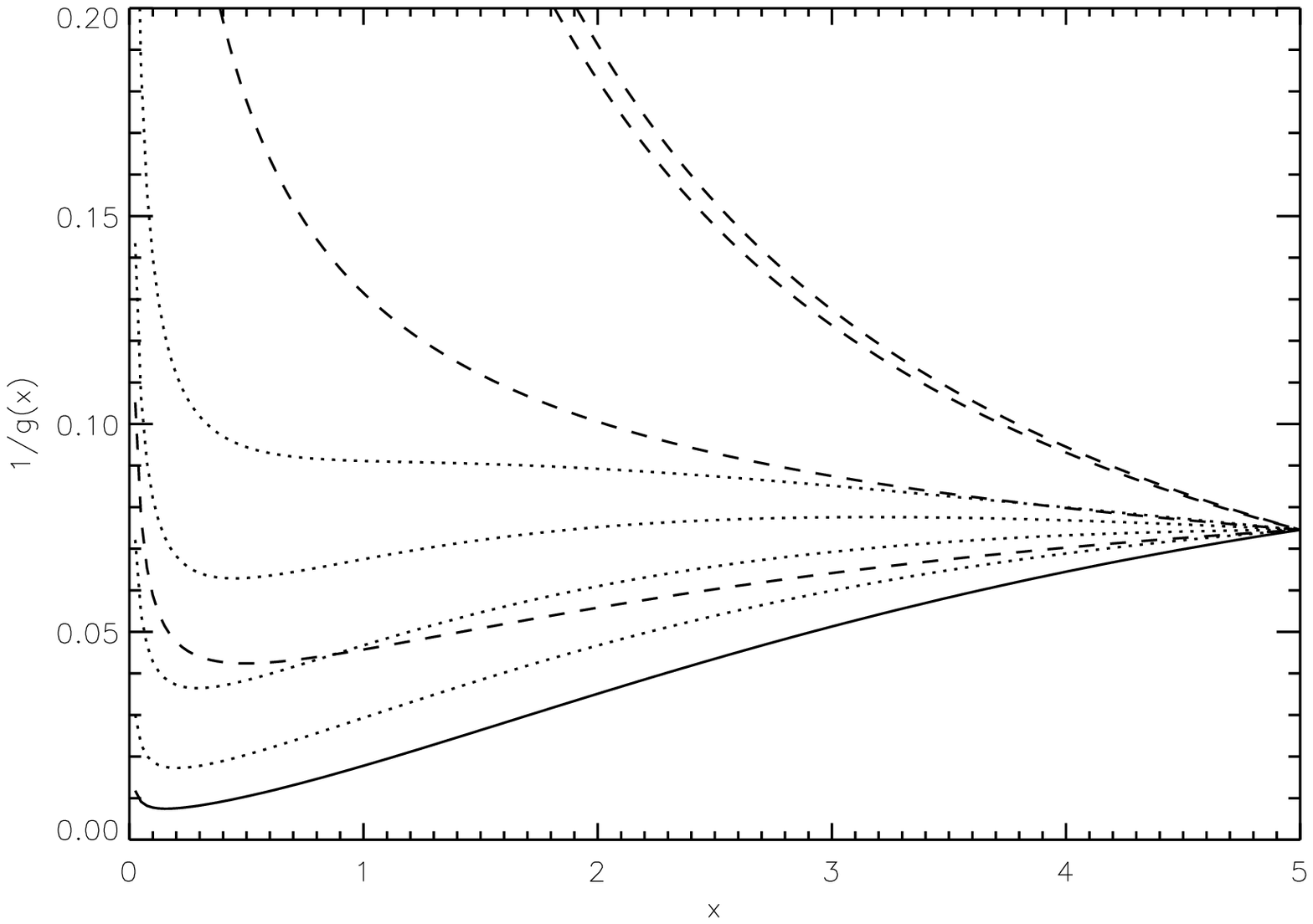,width=\figwidth}}
\caption[Examples of $1/g(x)$ ($x$ is the radius in a halo), which is
roughly proportional to $\tau(x)$, the ratio of cooling time to
free-fall time] {Plot of $1/g(x)$ (which is roughly proportional to
  $\tau(x)$), using the same parameters and linestyles as in
  Fig.~\ref{figrhorange}. The solid curve is given by $\gamma=1$ and
  $\eta_{200}=10$; the dotted curves are obtained by reducing
  $\eta_{200}$, as in Model B, and the dashed curves by increasing
  $\gamma$, as in Model A (see text for discussion).}
\label{figgofx}
\end{figure}

In Fig.~\ref{figgofx} we illustrate the general behaviour of $1/g(x)$,
using $c=5$ and the same values as in Fig.~\ref{figrhorange} for
$\gamma$ and $\eta_{200}$.  The qualitative behaviour is the same for
other values of $c$. For small enough $x$, $1/g(x)$ always diverges.
This is simply due to the divergence of the NFW density profile and as
long as the minimum occurs at sufficiently small $x$, as with the
solid curve ($\gamma=1$ and $\eta_{200}=10$), it is ignored. As we
decrease $\eta_{200}$ or increase $\gamma$, the minimum moves to larger
radii and $1/g(x)$ becomes a flatter function. Eventually, the minimum
disappears and if $\gamma$ is large, $1/g(x)$ becomes a steep
decreasing function of $x$. 

Since $\tau$ and its interpretation are approximate, we use an
algorithm which is relatively simple. The criteria for whether
$1/g(x)$ is well-behaved is given by its slope at $x=0.5$. When this
slope is positive (the well-behaved case), $1/g(x)$ is sure to have a
minimum inside $x=0.5$. If $\tau>\tau_0$ at this minimum, then $x_{\rm
  cf}=0$; if $\tau(c)<\tau_0$, then $x_{\rm cf}=c$; otherwise,
$\tau(x_{\rm cf})=\tau_0$ is solved numerically. 

When this slope is negative instead, $1/g(x)$ is either relatively
flat or strongly decreasing at larger radii (see Fig.~\ref{figgofx}; note
that the latter occurs in Model A but not Model
B). In this case, if $\tau(0.5)>\tau_0$ ($x=0.5$ being where the slope
is measured), then $x_{\rm cf}=0$ and all the gas is considered hot.
We have made the assumption that even if $\tau<\tau_0$ at larger radii,
there is sufficient hot gas in the centre to provide a working surface
on which infalling gas can shock to high temperatures, so that a
hydrostatic atmosphere can still form (as discussed in
section~\ref{briefsec}). If $\tau(0.5)<\tau_0$, then $x_{\rm cf}=c$
and all the gas is considered cold. (For completeness, the algorithm
actually allows for the situation, very rare in Models A and B, when
$\tau(c)>\tau(0.5)$ in a `poorly-behaved' halo. In this case, it finds
a numerical solution if $\tau_0$ lies between $\tau(0.5)$ and
$\tau(c)$.)

\subsection{The fraction, $f_{\rm unbind}$, of cold gas that forms stars}
Supernova feedback from star formation is assumed to eject the rest of the
gas once there is sufficient energy to do so. The fraction, $f_{\rm
  unbind}$, of cold gas that forms stars is given by,
\begin{eqnarray}
  \lefteqn{f_{\rm unbind}{M_{\rm gas}(x<x_{\rm cf})\over M_{\rm SN}}
    \epsilon_{\rm SN} 4\times 10^{50}{\rm erg}
       =M_{\rm gas}(x>x_{\rm cf})\, |E_{\rm gas}| }\nonumber\\
&&\hspace{0.5in}
  +\,(1-f_{\rm unbind})M_{\rm gas}(x<x_{\rm cf})\, |E_{\rm cold}|,
\end{eqnarray}
where $M_{\rm gas}$ is the total gas mass in the specified region and
$M_{\rm SN}$ is the mass of stars formed per resulting Type II
supernova. For a standard IMF $M_{\rm SN}=80$ \Ms\ \cite{tf90}. Since
we boost supernova rates by a factor of 5, $M_{\rm SN}=16$ \Ms\ in
this paper. The energy released by one supernova into surrounding gas
is $\epsilon_{\rm SN}\; 4\times 10^{50}$ erg. $E_{\rm gas}$ is defined
in equation~\ref{egas} and $E_{\rm cold}$ is defined as for $E_{\rm
  gas}$ except that the thermal term is set equal to zero. $|E_{\rm
  gas}|$ and $|E_{\rm cold}|$ are average quantities which estimate
the energies per unit mass required to eject the hot and cold gas
respectively.

If the solution to $f_{\rm unbind}$ in the above equation is greater
than 1, then the gas halo is not ejected. In this case all the cold
gas is able to form stars and $f_{\rm unbind}=1$.

\subsection{The mass of BDM that forms from hot gas}

Whenever there is hot gas in a halo, some of it may be able to cool to
form baryonic dark matter (BDM) before the next collapse.
The cooling radius, $r_{\rm cool}$, is obtained by solving numerically
the equation:
\begin{equation}\label{rcool}
  \left.{3\over 2}{\rho_{\rm g} 
     kT/\mu m_{\rm H}\over n_e n_{\rm H} \Lambda(T)}\right|_{r=r_{\rm cool}}
                 = \Delta t,
\end{equation}
where the lhs is the cooling time and $\Delta t$ is the time from
virialization to the next collapse or the present day, whichever
is sooner.

The mass of BDM formed is equal to the mass of gas inside $r_{\rm
  cool}$ {\rm minus} the mass which has already formed stars, if any.
Sometimes no hot gas is able to cool in the given time, in which case
no cooling flow operates.

\subsection{The mass cooling rate, $\dot{M}$}

The instantaneous mass cooling rate, $\dot{M}$, is estimated using
\begin{equation}
  \dot{M}=\left.{\d M_{\rm gas}(r)\over \d r}\right|_{r=r_{\rm cool}}
          \left.{\d r_{\rm cool}(t)\over \d t}\right|_{t=\Delta t},
\end{equation}
where $M_{\rm gas}(r)$ is the gas mass inside a radius of $r$, $t$ is
the time since the virialization and $\Delta t$ is as defined above.
The cooling radius as a function of $t$, $r_{\rm cool}(t)$, is
obtained by substituting $t$ for $\Delta t$ in equation~\ref{rcool}.

By differentiating equation~\ref{rcool} with
respect to $r_{\rm cool}$, we obtain
\begin{equation}
  {\d t\over\d r_{\rm cool}(t)} \approx
    {3\over 2}{\rho_{\rm g}^2\over n_e n_{\rm H}}
    {\alpha\over\eta_{200}\rho_{\rm g,200}r_s \Lambda(T)} 
  {\d\over\d x}\left(T\rho_{\rm g,200}\over T_{200}\rho_{\rm g}\right),
\end{equation}
where we have assumed that $\d\Lambda(T)/\d r$ is small.
Expanding the derivative gives
\begin{equation}
  {\d\over\d x}\left(T\rho_{\rm g,200}\over T_{200}\rho_{\rm
  g}\right)=
    {\gamma-2\over\gamma}\eta_{200}{\rho_{\rm g,200}\over\rho_{\rm g}}
    \left({1\over x(1+x)}-{\ln(1+x)\over x^2}\right).
\end{equation}

Since $\d M_{\rm gas}(r)/\d r=4\pi\rho_{\rm g}(r)r^2$, it follows that
\begin{equation}
  \dot{M}=4\pi{2\over 3}{n_e n_{\rm H}\over\rho_{\rm g}^2}
          {r_s^3\rho_{\rm g,200}^2\over\alpha}{\gamma\over\gamma-2}
    \left.{x^2 (\rho_{\rm g}/\rho_{\rm g,200})^2 \Lambda(T) \over
    \left({1\over x(1+x)}-{\ln(1+x)\over x^2}\right)}\right|_{x=x_{\rm cool}}.
\end{equation}

\subsection{The cooling flow power}

The cooling flow power is the bolometric luminosity of the cooling
flow region. It is given by
\begin{equation}
  {5\over 2}{\dot{M}k\,T(r=r_{\rm cool}) \over\mu m_{\rm H}},
\end{equation}
which uses the enthalpy, $5 k T/(2\mu m_{\rm H})$, to
estimate the total energy radiated per unit mass.
It corresponds observationally to the bolometric luminosity
inside the cooling radius.

\subsection{The X-ray luminosity, $L_{\rm X}$}

This is the sum of the cooling flow power and the bolometric
luminosity due to the gas outside $r_{\rm cool}$:
\begin{equation} \label{eqlx}
  L_{\rm X}=\int_{r_{\rm cool}}^{r_{200}}\!\!\!\!\!\!
    n_e n_{\rm H}\Lambda(T)4\pi r^2\d r \;+\;
             {5\over 2}{\dot{M}k\,T(r=r_{\rm cool}) \over\mu m_{\rm H}}.
\end{equation}

By the time of observation, the density profile of gas that belonged
to $r<r_{\rm cool}$ differs substantially from that of the notional
gas profile due to the effects of radiative cooling. Hence the cooling
flow power is estimated separately. Although the changes due to
cooling are also felt outside $r_{\rm cool}$, because volume is a
rapidly increasing function of the radius, the effect is only
significant close to $r_{\rm cool}$, so that we treat the atmosphere
as unmodified from the notional gas profile outside $r_{\rm cool}$.

\subsection{The emission-weighted temperature}

This is the temperature that is implied whenever we refer to the
temperature of a cluster as a whole (as in section~\ref{clustersec}).
We calculate the temperature as weighted by the luminosity outside
$r_{\rm cool}$. It is thus given by
\begin{equation}
  T_{\rm ew}={\int_{r_{\rm cool}}^{r_{200}}
                       T(r)n_e n_{\rm H}\Lambda(T)4\pi r^2\d r
          \over \int_{r_{\rm cool}}^{r_{200}}
                 n_e n_{\rm H}\Lambda(T)4\pi r^2\d r}.
\end{equation}

\section{Gas energy equation}

We derive below the equations that govern $E_{\rm gas}$ and $e_{\rm
  gas}$.  $E_{\rm gas}$, defined in equation~\ref{egasg}, is the
mean total specific energy of gas in a proto-halo. The definitions of
proto-halo and proto-gas halo are given in section~\ref{ensec}. We
use $e_{\rm gas}$ (equation~\ref{eegas}) to follow the total specific
energy of the gas at a position which moves with the gas.

The gas equations may be written
\begin{equation}\label{masscons}
\dpar{\rhog}{t} + \nabla \cdot \rhog \vel = 0,
\end{equation}
for the conservation of mass,
\begin{equation}\label{navstok}
\rhog {d\vel\over dt} = - \nabla P  + \nabla \cdot \stress 
- \rhog \nabla \phi,
\end{equation}
for the conservation of momentum and 
\begin{equation}\label{enent}
\rhog T {dS\over dt} = \sum_{i,j=1}^3 T_{ij} \dpar{v_i}{x_j} + \Gamma,
\end{equation}
for the conservation of energy.  Here $\rhog$, $T$, $P$, $S$ and
$\vel$ are, respectively, the density, temperature, pressure, specific
entropy and velocity of the gas, $\phi$ is the gravitational
potential, and $\stress$ the viscous stress tensor (with components
$T_{ij}$).  The Lagrangian time derivative is
\begin{equation} \label{lagrange}
{d\over dt} = \dpar{}{t} + \vel \cdot \nabla.
\end{equation}
The first term on the right in the energy equation is the viscous
heating rate.  The second term, $\Gamma$, is the net additional
heating rate per unit volume due to effects other than adiabatic and
viscous heating.  Such processes include supernova heating and
radiative heat loss.

The specific enthalpy is defined as $H = \epsilon + P V$, where
$\epsilon$ is the specific energy and $V = 1/\rhog$ is the specific
volume.  Using the first law of thermodynamics, $d\epsilon = T\, dS -
P \, dV$, gives $ dH = T\, dS + V \, dP$, so that 
\begin{equation}
\rhog {dH\over dt} = \rhog T {dS\over dt} + {dP\over dt}
= \sum_{i,j=1}^3 T_{ij} \dpar{v_i}{x_j} + \Gamma + \dpar{P}{t} +
\vel \cdot \nabla P,
\end{equation}
where we have used the energy equation (\ref{enent}) and expanded the
Lagrangian derivative.  Using the momentum equation (\ref{navstok}) to
replace $\nabla P$ in the last term gives, after some algebra,
\begin{equation}
\rhog {dH\over dt} = \nabla \cdot (\stress\vel) + \Gamma + \dpar{P}{t}
- \rhog {d\over dt} \half\vel^2 - \rhog\vel \cdot \nabla\phi.
\end{equation}
Converting $\vel \cdot \nabla\phi$ in the last term into time
derivatives of $\phi$, and rearranging, we get
\begin{equation}
\rho_{\rm g} {d\over dt} \left( H + \half {\bf v}^2 + \phi \right)
- \dpar{P}{t} = \rho_{\rm g} \dpar{\phi}{t} + \nabla \cdot ({\bf T v})
+ \Gamma.
\end{equation}
Using equation~\ref{masscons} and $\rho_{\rm g} H - P = \rho_{\rm g}
\epsilon$, this can be rewritten as
\begin{eqnarray}
\lefteqn{\dpar{}{t} \left[ \rho_{\rm g} \left(\epsilon + \half {\bf v}^2 +
\phi \right) \right] + \nabla \cdot \left[ \rho_{\rm g} {\bf v} 
\left(H + \half {\bf v}^2 + \phi \right)\right] = }\nonumber\\
&&  \rho_{\rm g}\dpar{\phi}{t} + \nabla \cdot ({\bf T v}) + \Gamma.
\end{eqnarray}
Integrating this over a comoving volume $V$, we get
\begin{eqnarray} \label{longB}
\lefteqn{ \int_V  \dpar{}{t}\left[\rho_{\rm g} 
\left(\epsilon + \half {\bf v}^2 + \phi \right)\right] \, \d V = }\nonumber\\
&& \int_{\partial V} \left[ -\rho_{\rm g} {\bf v}
\left(H + \half {\bf v}^2 + \phi\right) + {\bf Tv} \right] \cdot \d {\bf A} 
\nonumber\\
&& + \int_V \left(\rho_{\rm g} \dpar{\phi}{t} + \Gamma\right) \, \d V,
\end{eqnarray}
where $\d {\bf A}$ is a vector element of surface area.
But, for any $Q$ and comoving volume $V$,
\begin{equation}
{\d\over\d t} \int_V Q \, \d V = \int_V \dpar{}{t} Q \, \d V +
\int_{\partial V} Q {\bf v} \cdot \d{\bf A},
\end{equation}
so that when the partial derivative on the lhs of equation~\ref{longB}
is taken outside the integral, we get extra terms
which cancel most of the surface terms, giving
\begin{eqnarray} \label{B2}
\lefteqn{ {\d\over\d t} \int_V  \left[\rho_{\rm g} 
\left(\epsilon + \half {\bf v}^2 + \phi \right)\right] \, \d V =}\nonumber\\
&& \int (-P\vel+\stress\vel) \cdot\d{\bf A}
   + \int_V \left( \rhog \dpar{\phi}{t} + \Gamma \right) \,\d V.
\end{eqnarray}

In section~\ref{mech1} we assume a simplified scenario where no gas is
`removed' to form stars and BDM in the proto-halo.  We also assume
that the gas pressure and viscosity at the boundary of the proto-halo
are negligible. If $V$ is the volume occupied by gas in the proto-halo,
then the surface integral above vanishes. 
Substituting $3kT/(2\mu m_{\rm H})$ for $\epsilon$ and using the
definition of $E_{\rm gas}$ (equation~\ref{egasg}), we obtain the result
\begin{equation} \label{B1}
  {\d E_{\rm gas} \over\d t} = {1\over M_{\rm gas}}
\int_V \left(\rho_{\rm g} \dpar{\phi}{t} + \Gamma \right) \, \d V.
\end{equation}

In sections~\ref{mech2} to \ref{mech4}, we follow the gas in the
proto-halo in more detail, defining the proto-gas halo to include only
gas that eventually belongs to the virialized gas halo. Thus the mass
of the proto-gas halo is constant with time.  The volume, $V$, that it occupies
is irregular at early times, containing `pockets' of gas which are
excluded from the proto-gas halo because they later convert into stars
or BDM. Using $e_{\rm gas}= (\epsilon + \half {\bf v}^2 + \phi)$ and
$\d m=\rho_{\rm gas} \d V$, we rewrite equation~\ref{B2} as
\begin{equation}
{\d\over\d t} \int e_{\rm gas} \,\d m
 = \int (-P\vel+\stress\vel) \cdot\d{\bf A}
   + \int_V \left( \rhog \dpar{\phi}{t} + \Gamma \right) \,\d V,
\end{equation}
where $\int e_{\rm gas} \,\d m$ is the total energy of the proto-gas
halo.  The surface integral gives the rate at which the proto-gas halo
does work on neighbouring gas.  In section~\ref{ensec} we investigate
the pressure term only.

\section{The evolution of $E_{\rm gas,G}$}

In this appendix, we obtain a simple expression for the variation of
$E_{\rm gas,G}$ with time (where the subscript `G' implies that the
system is evolved without including non-gravitational processes), and
obtain rough estimates of $E_{\rm gas,G}$ at $t_a$ and $t_v$ (see
Fig.~\ref{figensec}).

If the gas and dark matter have the same distribution, then $4 \pi G
\rho_{\rm g,G} = f_{\rm gas} \nabla^2 \phi $ for some constant $f_{\rm
  gas} < 1$. Now, by integrating by parts twice, we obtain Green's
  Theorem: 
\begin{equation}
  \int \phi\, \nabla^2\dot{\phi} \,\d V \equiv
   \int \left(\phi {\bf \nabla}\dot{\phi} - \dot{\phi} {\bf \nabla}\phi 
                                  \right)\cdot\d{\bf A}
       + \int \dot{\phi}\,\nabla^2\phi \,\d V,
\end{equation}
where $\dot{\phi}=\partial\phi/\partial t$. If the integrals are made
over all space and $\phi$ vanishes at infinity, then the surface
integrals vanish. Since $\rho_{\rm g,G}\propto \nabla^2 \phi $, it
follows that
\begin{equation}
  \int \rho_{\rm g,G} \dpar{\phi}{t} \, \d V
  = \dpar{}{t} \int \half \rho_{\rm g,G} \phi \, \d V.
\end{equation}
In order to substitute into equation~\ref{B1}, where the
integration is made over the volume of the proto-halo only, we need to
assume that $\rho_{\rm g,G}=\rho_{\rm tot}=0$ outside the proto-halo.
The volume of integration above can then be shrunk down to the
proto-halo. Setting $\Gamma=0$, equation~\ref{B1} gives
\begin{equation}
  {\d E_{\rm gas,G}\over \d t} = {1\over M_{\rm gas}} {\d\over\d t}
                 \int \half \rho_{\rm g,G} \phi \, \d V.
\end{equation}
Therefore,
\begin{equation}
  E_{\rm gas,G} = {1\over M_{\rm gas}} 
    \int \half \rho_{\rm g,G} \phi \, \d V + \mbox{constant}.
\end{equation}

In section~\ref{mech1}, we define the quantity $E_{\rm max}=[E_{\rm
  gas,G}]^{t_a}_{t_v}$. The above result thus implies that
\begin{equation}
  E_{\rm max}={1\over M_{\rm gas}}
       \left[\int\half\rho_{\rm g,G}\phi\,\d V\right]_{t_v}^{t_a}
  \approx {1\over M_{\rm gas}} 
       \left.\int-{1\over 4}\rho_{\rm g,G}\phi\,\d V \right|_{t_v},
\end{equation}
where we have assumed that the integral scales as the
inverse of the radius of the system, and that the turnaround radius
is twice the virial radius.

To obtain a rough estimate of the {\em absolute} value of $E_{\rm gas,G}$ at
$t_v$, we assume that the kinetic term in equation~\ref{egasg} is
zero, and estimate the thermal term.  The gravitational binding energy
of the halo is equal to $\int(1/2)\rho_{\rm tot}\phi\,\d V$. The
virial theorem then implies that the thermal energy of the gas halo is
approximately $f_{\rm gas}(-1/2)\int(1/2)\rho_{\rm tot}\phi\,\d V$,
where $f_{\rm gas}=\rho_{\rm g,G}/\rho_{\rm tot}$ is a constant and
possible boundary terms at $r_{200}$ have been ignored.  Dividing by
$M_{\rm gas}$ gives the specific thermal energy of the gas:
$\int(-1/4)\rho_{\rm g,G}\phi\,\d V/M_{\rm gas}$. Therefore,
\begin{equation} \label{behalo}
  E_{\rm gas,G}(t_v)\approx {1\over M_{\rm gas}}
      \left.\int {3\over 4} \rho_{\rm g,G} \phi\, \d V\right|_{t_v}
   \approx -3 E_{\rm max},
\end{equation}
as shown in Fig.~\ref{figensec}.
It follows that $E_{\rm gas,G}(t_a)\approx -2 E_{\rm max}$.

\end{document}